\documentclass[journal,twoside,web]{ieeecolor}

\usepackage{T_AutomaticControl}
\usepackage{cite}
\usepackage{amsmath,amssymb,amsfonts}
\usepackage{algorithm,algorithmic}
\usepackage{cases}
\usepackage{graphicx}
\usepackage{textcomp}
\usepackage{dsfont}
\usepackage{enumerate}

\usepackage[caption=false,font=footnotesize]{subfig}

\usepackage{theorem}
\theorembodyfont{\rmfamily}

\newtheorem{lem}{Lemma}
\newtheorem{defin}{Definition}
\newtheorem{theorem}{Theorem}

\newtheorem{assum}{Assumption}

\renewcommand{\QED}{\QEDopen}

\newcommand{\NN}{\mathbb{N}}
\newcommand{\Zp}{{\mathbb{Z}_+}}
\newcommand{\mc}[1]{\mathcal{#1}}
\newcommand{\hs}{& \hspace{-3mm}}
\newcommand{\ssen}{s^{\rm s}}
\newcommand{\mcss}{\mathcal{S}^{\rm s}}
\newcommand{\hssen}{\hat{s}^{\rm s}}
\newcommand{\hsrec}{\hat{s}^{\rm r}}
\newcommand{\tssen}{\tilde{s}^{\rm s}}
\newcommand{\tsrec}{\tilde{s}^{\rm r}}

\newcommand{\srec}{s^{\rm r}}
\newcommand{\mcsr}{\mathcal{S}^{\rm r}}

\newcommand{\Us}{U^{\rm s}}
\newcommand{\Ur}{U^{\rm r}}
\newcommand{\bUs}{\bar{U}^{\rm s}}
\newcommand{\iTheta}{{\it \Theta}}
\newcommand{\io}{{\rm i.o.}}
\newcommand{\tb}{\theta_{\rm b}}
\newcommand{\tm}{\theta_{\rm m}}
\newcommand{\ssk}{s^{\rm s}_k}
\newcommand{\srk}{s^{\rm r}_k}

\newcommand{\Exp}{\mathbb{E}}
\newcommand{\Prob}{\mathbb{P}}

\newcommand{\iThetas}{{\it \Theta}^{\rm s}}
\newcommand{\iThetar}{{\it \Theta}^{\rm r}}
\newcommand{\pis}{\pi^{\rm s}}
\newcommand{\pir}{\pi^{\rm r}}
\newcommand{\hpis}{\hat{\pi}^{\rm s}}

\newcommand{\tsm}{\theta^{\rm s}_{\rm m}}
\newcommand{\tsb}{\theta^{\rm s}_{\rm b}}
\newcommand{\tru}{\theta^{\rm r}_{\rm u}}
\newcommand{\tra}{\theta^{\rm r}_{\rm a}}
\newcommand{\ts}{\theta^{\rm s}}
\newcommand{\tr}{\theta^{\rm r}}
\newcommand{\BRs}{{\rm BR}^{\rm s}}
\newcommand{\BRr}{{\rm BR}^{\rm r}}

\newcommand{\bss}{\bar{s}^{\rm s}}
\newcommand{\bsr}{\bar{s}^{\rm r}}
\newcommand{\Ts}{\Theta^{\rm s}}
\newcommand{\Tr}{\Theta^{\rm r}}

\DeclareMathOperator*{\argmax}{arg\,max}

\def\TitleSSHR{Asymptotic Security using Bayesian Defense Mechanism with Application to Cyber Deception}

\usepackage{etoolbox}
\makeatletter
\@ifundefined{color@begingroup}%
  {\let\color@begingroup\relax
   \let\color@endgroup\relax}{}%
\def\fix@ieeecolor@hbox#1{%
  \hbox{\color@begingroup#1\color@endgroup}}
\patchcmd\@makecaption{\hbox}{\fix@ieeecolor@hbox}{}{\FAILED}
\patchcmd\@makecaption{\hbox}{\fix@ieeecolor@hbox}{}{\FAILED}
\begin{document}
\title{\TitleSSHR}
\author{Hampei Sasahara, \IEEEmembership{Member, IEEE}, Henrik Sandberg, \IEEEmembership{Senior Member, IEEE}
\thanks{H.~Sasahara is with the Department of Systems and Control Engineering, Tokyo Institute of Technology, Tokyo, 152-8552 Japan e-mail: sasahara@sc.e.titech.ac.jp.}
\thanks{H.~Sandberg is with the Division of Decision and Control Systems, KTH Royal Institute of Technology, Stockholm, SE-100 44 Sweden e-mail: hsan@kth.se.}
\thanks{This work was supported in part by JSPS KAKENHI Grant Number 22K21272, the Swedish Research Council grant 2016-00861, and Digital Futures (projects DEMOCRITUS and RoSE).}
\thanks{\copyright 2023 IEEE. Personal use of this material is permitted. Permission from IEEE must be
obtained for all other uses, in any current or future media, including
reprinting/republishing this material for advertising or promotional purposes, creating new
collective works, for resale or redistribution to servers or lists, or reuse of any copyrighted
component of this work in other works.}}

\maketitle

\begin{abstract}
This paper addresses the question whether model knowledge can guide a defender to appropriate decisions, or not, when an attacker intrudes into control systems.
The model-based defense scheme considered in this study, namely Bayesian defense mechanism, chooses reasonable reactions through observation of the system's behavior using models of the system's stochastic dynamics, the vulnerability to be exploited, and the attacker's objective.
On the other hand, rational attackers take deceptive strategies for misleading the defender into making inappropriate decisions.
In this paper, their dynamic decision making is formulated as a stochastic signaling game.
It is shown that the belief of the true scenario has a limit in a stochastic sense at an equilibrium based on martingale analysis.
This fact implies that there are only two possible cases: the defender asymptotically detects the attack with a firm belief, or the attacker takes actions such that the system's behavior becomes nominal after a finite number of time steps.
Consequently, if different scenarios result in different stochastic behaviors, the Bayesian defense mechanism guarantees the system to be secure in an asymptotic manner provided that effective countermeasures are implemented.
As an application of the finding, a defensive deception utilizing asymmetric recognition of vulnerabilities exploited by the attacker is analyzed.
It is shown that the attacker possibly withdraws even if the defender is unaware of the exploited vulnerabilities, as long as the defender's unawareness is concealed by the defensive deception.
\end{abstract}

\begin{IEEEkeywords}
Bayesian methods, game theory, intrusion detection, security, stochastic systems.
\end{IEEEkeywords}

\section{Introduction}
\label{sec:introduction}

\IEEEPARstart{S}{ocietal} monetary loss from cyber crime is estimated to be about a thousand billion USD per year presently, and even worse, a rising trend can be observed~\cite{Mcafee2020The}.
Another trend is that not only information systems but also control systems, which are typically governed by physical laws, are exposed to cyber threats as demonstrated by recent incidents~\cite{CISA2014,CISA2017,CISA2018,CISA2021}.
\emph{Deception} is a key notion to predict the consequence of incidents.
Rational attackers take deceptive strategies, i.e., the attacker tries to conceal her existence and even mislead the defender into taking inappropriate decisions.
An example of deception is replay attacks, which hijacks sensors of the plant, eavesdrops the nominal data transmitted when the system is operated under normal conditions, and replays the observed nominal data during the execution of another damaging attack.
A replay attack was executed in the Stuxnet incident, and it was an essential factor leading to serious damage in the targeted plant~\cite{Nicolas2011Stuxnet}.
The incident suggests that prevention of deception is a fundamental requirement for secure system design.

Assuming the situation where an attacker might intrude into a control system where a defense mechanism is implemented, this paper addresses the following question:
\emph{Can model knowledge guide the defender to appropriate decisions against attacker's deceptive strategies?}
Specifically, we consider the case where the stochastic model of the control system, the vulnerability to be exploited, and the objective of the attacker are known.
The setting naturally leads to \emph{Bayesian defense mechanisms}, which monitor the system's behavior and form a belief on the existence of the attacker using the model.
If the system's behavior is inconsistent with the nominal one, the belief increases owing to Bayes' rule.
When the belief is strong enough, the Bayesian defense mechanism proactively carries out a proper reaction.
On the other hand, we also suppose a powerful attacker who knows the model and the defense scheme to be implemented.
The attacker aims at achieving her objective while avoiding being detected by deceiving the defender.

For mathematical analysis, we formulate the decision making as a dynamic game with incomplete information.
More specifically, we refer to the game as a stochastic signaling game, because it is a stochastic game~\cite{Neyman2003Stochastic} in the sense that the system's dynamics is given as a Markov decision process (MDP) governed by two players and it is also a signaling game~\cite{Cho1987Signaling} in the sense that one player's type is unknown to the opponent.
In this game, the attacker strategically chooses harmful actions while avoiding being detected, while the defender, namely, the Bayesian defense mechanism, chooses appropriate counteractions according to her belief.

Based on the game-theoretic formulation, we find that \emph{model knowledge can always lead the defender to appropriate decisions in an asymptotic sense} as long as the system's dynamics admits no stealthy attacks.
More specifically, there are only two possible cases: one is that the defender asymptotically forms a firm belief on the existence of an attacker and the other is that the attacker takes harmless actions after finite time such that the system converges to nominal behavior.
This finding leads to the conclusion that the Bayesian defense mechanism guarantees the system to be secure in an asymptotic manner.

The analysis means that the defender always wins in an asymptotic manner when the stochastic model of the system is available and \emph{the vulnerability exploited for the intrusion is known and modeled.}
However, in practice, it is hard to be aware of all possible vulnerabilities in advance.
As an application of the finding above, we consider \emph{defensive deception using bluffing} that utilizes asymmetric recognitions between the attacker and the defender.
Specifically, we suppose that, the defender is unaware of the exploited vulnerability but the attacker is unaware of the defender's unawareness.
If the state of the system does not possess any information about the defender's recognition on the vulnerability, the attacker cannot identify whether the defender is aware of the vulnerability, or not.
The result obtained in the former part suggests that the attacker may possibly withdraw
if the defender's reactions affect only the attacker's utility without influence to the system's behavior.
The difficulty of the analysis is that standard incomplete information games, which assume common prior, cannot describe this situation.
The common prior implicitly assumes that the attacker is aware of the defender's unawareness.
To overcome the difficulty, we employ the Mertens-Zamir model, which can represent incomplete information games without common prior assumption, using the notion of belief hierarchy~\cite{Mertens1985Formulation,Dekel2015Epistemic}.
Based on this setting, we show, in a formal manner, that the defensive deception effectively works when the attacker strongly believes that the defender is aware of the vulnerability.

\subsection*{Related Work}

Model-based security analysis helps the system designer to prioritize security investments~\cite{Lund2011Model}.
Attack graphs~\cite{Philips1998A} and attack trees~\cite{Scheneier1994Attack} are basic models of vulnerabilities, attacks, and consequences.
Incorporating defensive actions into the graphical representation induces defense trees~\cite{Bistarelli2006Defense}.
For dynamic models, attack countermeasure trees, partially observable MDP, and Bayesian network model have been used~\cite{Roy2012Attack,Miehling2018A,Chockalingam2017Bayesian}.
Those probabilistic models naturally lead to Bayesian defense mechanisms, such as Bayesian intrusion detection~\cite{Kruegel2003Bayesian,Alhakami2019Network}, Bayesian intrusion response~\cite{Zonouz2014RRE}, and Bayesian security risk management~\cite{Poolsappasit2012Dynamic}.
Meanwhile, the model of the dynamical system to be protected is also used for control system security~\cite{Sandberg2015Cyberphysical,Seyed2019Systems}.
For example, identifying existence of stealthy attacks and removing the vulnerability require the dynamical model~\cite{Teixeira2012Revealing,Milosevic2020Security}, and attack detection performance can be enhanced by model knowledge~\cite{Giraldo2018A}.
Our Bayesian defense mechanisms can be interpreted as a generalization of those approaches.
This work reveals a fundamental property of such commonly used model-based defense schemes.



Game theory is a standard approach to modeling the decision making in cyber security, where there inevitably arises a need to address strategic interactions between the attacker and the defender~\cite{Tambe2012Security,Alpcan2010Network}.
In particular, games with incomplete information play a crucial role in deceptive situations~\cite{Pawlick2019Taxonomy,Sayin2021Deception,Sasahara2021Epistemic,Pawlick2021Game}.
The modeling in this study follows the signaling game framework in~\cite{Pawlick2019Modeling,Zhu2020Secure}.
Our main concern is especially on asymptotic phenomena in the dynamic deception and effectiveness of model knowledge.

Our finding is based on analysis of an asymptotic behavior of Bayesian inference.
The convergence property of Bayesian inference on the true parameter is referred to as Bayesian consistency, which has been investigated mainly in the context of statistics~\cite{Diaconis1986On,Walker2004New}.
However, those existing results are basically applicable only to independent and identically distributed (i.i.d.) samples because the discussion mostly relies on the strong law of large numbers (SLLN).
Although there is an extension to Markov chains~\cite{Peter2002Bayesian}, the observable variable in our work is not Markov.
Indeed, sophisticated attackers can choose strategies such that the states at all steps are correlated with the entire previous trajectory.
Thus, existing results for Bayesian consistency cannot be applied to our problem in a straightforward manner.

Preliminary versions of this work have been presented in~\cite{Sasahara2020Asymptotic,Sasahara2021Asymptotic}, but they made the claim of Theorem~\ref{thm:b_zero} as an assumption rather than proving it.
Moreover, they did not include rigorous proofs of the claims in Section~\ref{sec:analysis} and analysis of the bluffing proposed in Section~\ref{sec:app}.

\subsection*{Organization and Preliminaries}

In Section~\ref{sec:modeling}, we present a motivating example of water supply networks, and subsequently, formulate the decision making as a stochastic signaling game.
Section~\ref{sec:analysis} analyzes the consequence of the formulated game and shows that Bayesian defense mechanisms can achieve asymptotic security of the system to be protected.
In Section~\ref{sec:app}, we analyze a defensive deception that utilizes asymmetric recognition as an application of the finding of Section~\ref{sec:analysis}.
The game of interest is reformulated using the Mertens-Zamir model.
It is shown that the attacker possibly stops the execution even if the defender is unaware of the exploited vulnerabilities, as long as the defender's belief is concealed.
Section~\ref{sec:sim} verifies the theoretical results through numerical simulation.
Finally, Section~\ref{sec:conc} concludes and summarizes the paper.

Let $\NN$, $\Zp$, and $\mathbb{R}$ be the sets of natural numbers, nonnegative integers, and real numbers, respectively.
The $k$-ary Cartesian power of the set $\mc{X}$ is denoted by $\mc{X}^k.$
The tuple $(x_0,\ldots,x_k)$ is denoted by $x_{0:k}$.
The cardinality of a set $\mc{X}$ is denoted by $|\mc{X}|$.
For a set $\mc{X},$ the Kronecker delta denoted by $\delta:\mc{X}\times\mc{X}\to \{0,1\}$ is defined by $\delta(x,y)=1$ if $x=y$ and $\delta(x,y)=0$ otherwise.
The $\sigma$-algebra generated by a random variable $X$ is denoted by $\sigma(X)$.
For a sequence of events $E_k$ for $k\in\NN$, the supremum set $\cap_{N=1}^{\infty}\cup_{k=N}^{\infty}E_k$, namely, the event where $E_k$ occurs infinitely often, is denoted by $\{E_k\ \io\}$.
Jensen's inequality, which is often applied in this paper, is given as follows: For a real convex function $\varphi$ and a finite set $\mc{X}$, the inequality
\begin{equation}\label{eq:Jensen}
 \textstyle{ \sum_{x\in\mc{X}} p(x)\varphi(a(x)) \geq \varphi\left(\sum_{x\in\mc{X}} p(x) a(x) \right)}
\end{equation}
holds where $a:\mc{X}\to\mathbb{R}$ and $p:\mc{X}\to[0,1]$ that satisfies the equation $\sum_{x\in\mc{X}}p(x)=1$.
The inequality is reversed if $\varphi$ is concave.
The generalized Borel-Cantelli's second lemma is given as follows~\cite[Theorem~4.3.4]{Durrett2019Probability}:
Let $\mc{F}_k$ for $k\in\Zp$ be a filtration of a probability space $(\Omega,\mc{F},\Prob)$ with $\mc{F}_0:=\{\emptyset,\Omega\}$ and let $E_k$ for $k\in\Zp$ be a sequence of events with $E_k\in\mc{F}_{k+1}$.
Then
\begin{equation}\label{eq:Borel-Cantelli}
\textstyle{
 \{E_k\ \io\} = \left\{
 \omega\in\Omega:
 \sum_{k=0}^\infty \Prob(E_k|\mc{F}_k)(\omega)=\infty
 \right\}.
}
\end{equation}
The appendix contains the proofs of the claims made in the paper.

\section{Modeling using Stochastic Signaling Games}
\label{sec:modeling}

\subsection{Motivating Example}
\label{subsec:sketch}

\begin{figure}[t]
  \centering
  \includegraphics[width=0.98\linewidth]{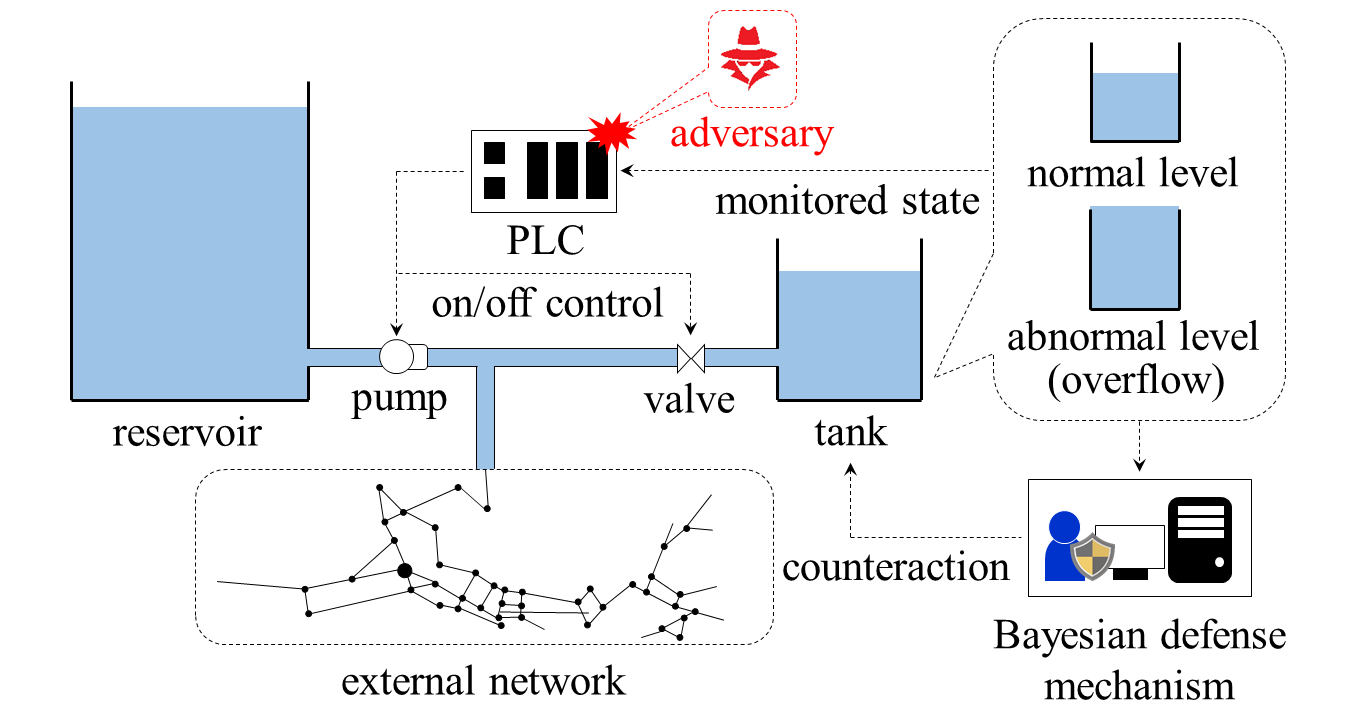}
  \caption{Motivating example: water tank system connected to a reservoir within a water distribution network.
  The programmable logic controller (PLC) transmits on/off control signals to the pump and the valve monitoring the state, the water level of the tank.
  In the scenario, an adversarial software possibly intrudes into the PLC and then the infected PLC tries to cause overflow by sending inappropriate control signals without being detected.
  A Bayesian defense mechanism, which utilizes the data of the monitored state and forms her belief on existence of the attacker based on the system model, is also equipped to deal with the attack.
  }
  \label{fig:WDN}
\end{figure}

As a motivating example, we consider water distribution networks (WDNs), which supply drinking water of suitable quality to customers.
Because of their indispensability to our life, WDNs are an attractive target for adversaries and expose their architecture to cyber-physical attacks~\cite{Rasekh2016Smart}.
In particular, we treat the water tank system illustrated by Fig.~\ref{fig:WDN}, where a tank is connected to a reservoir within a WDN.
The amount of the water in the tank varies due to usage for drinking and flow between the external network.
Thus the tank system is needed to be properly controlled through actuation of the pump and the valve to keep the water amount within a desired range~\cite{Creaco2019Real}.
A programmable logic controller (PLC) transmits on/off control signals to the pump and the valve monitoring the state, namely, the water level of the tank.
The dynamics is modeled as a MDP, where the state space and the action space are given by quantized water levels and finite control actions.
Interaction to the external network is modeled as the randomness in the process.

We here suppose an attack scenario considered in~\cite{Taormina2017Characterizing}.
The adversary succeeds to hijack the PLC and can directly manipulate its control logic.
Such an intrusion can be carried out by stealthy and evasive maneuvers in advanced persistent threats~\cite{Chen2014Study}.
The objective of the attack is to damage the system by causing water overflow through inappropriate control signals without being detected.
To deal with this attack, we consider a Bayesian defense mechanism, which utilizes the data of the monitored state and forms her belief on existence of the attacker based on the system model.
The Bayesian defense mechanism chooses a proper reaction by identifying if the system is under attack through an observation of the state.
If the system's behavior is highly suspicious, for example, the defense mechanism takes an aggressive reaction such as log analysis, dispatch of operators, or emergency shutdown.

The defender's belief on the existence of an attacker plays a key role to analyze the consequence of the threat.
When the attacker naively executes an attack, the system's behavior becomes different from the one of the normal operation and accordingly the belief increases.
On the other hand, if the attacker chooses sophisticated attacks that deceive the defender, the belief may decrease.
Our main interest in this study is to investigate the defense capability achieved by the Bayesian defense mechanism.

\subsection{Modeling using Stochastic Signaling Game}

We introduce the general description based on dynamic games with incomplete information.
In particular, we refer to the game as a stochastic signaling game where the system's dynamics is given as an MDP and the type of a player is unknown to the opponent.

\begin{figure}[t]
\centering
\includegraphics[width = .98\linewidth]{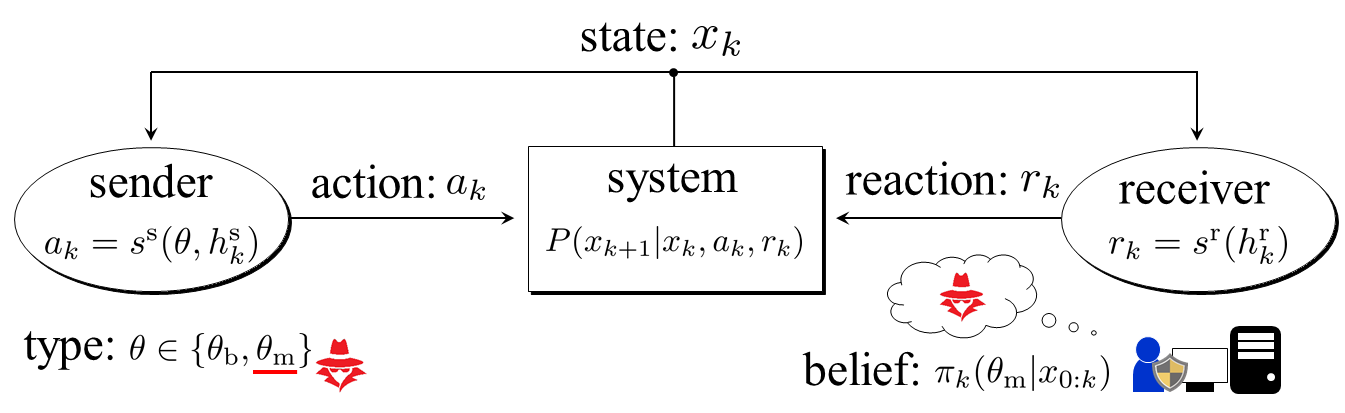}
\caption{
Block diagram of the system to be protected using the Bayesian defense mechanism.
The system is governed by actions and reactions, which are decided by the sender and the receiver, respectively.
The sender type $\tb$ means that the system is normally operated.
The other type $\tm$ means that there exists an attacker who executes malicious actions.
The receiver is the Bayesian defense mechanism that forms her belief on the existence of an attacker utilizing the measured data and chooses reactions based on the belief.
}
\label{fig:sys}
\end{figure}

The system to be protected with a Bayesian defense mechanism is depicted in Fig.~\ref{fig:sys}.
The system is modeled by a finite MDP governed by two players as in standard stochastic games.
Formally, the MDP considered in this paper is given by the tuple $\mc{M}:=(\mc{X},\mc{A},\mc{R},P,P_0)$ where $\mc{X}$ is a finite state space, $\mc{A}$ and $\mc{R}$ are finite action spaces, $P:\mc{X}\times\mc{X}\times\mc{A}\times\mc{R}\to[0,1]$ is a transition probability, and $P_0:\mc{X}\to[0,1]$ is the probability distribution of the initial state.
The state at the $k$th step is denoted by $x_k\in\mc{X}$.
There is an agent who can alter the system through an \emph{action} $a_k\in \mc{A}$ for $k\in \Zp$.
We refer to the agent as \emph{sender} as in standard signaling games.
Based on the measured output, the Bayesian defense mechanism, called a \emph{receiver}, chooses an action $r_k\in\mc{R}$ at each time step.
We henceforth refer to $r_k$ as a \emph{reaction} for emphasizing that $r_k$ denotes a counteraction against potentially malicious attacks.
The system dynamics is given by $P$, where the transition probability from $x$ to $x'$ with $a$ and $r$ is denoted by $P(x'|x,a,r)$.
To eliminate the possibility of trivial stealthy attacks, we assume that the system's behavior varies in a stochastic sense when different actions are taken.
\begin{assum}\label{assum:input_obs}
For any $x\in\mc{X}$ and $r\in\mc{R}$, there exists $x'\in\mc{X}$ such that
\begin{equation}\label{eq:dummy}
 P(x'|x,a,r)\neq P(x'|x,a',r)
\end{equation}
for different actions $a\neq a'$.
\end{assum}



Next, we determine the class of the decision rules.
Let $\theta\in\iTheta$ denote the \emph{type} of the sender.
For simplicity, the type is assumed to be binary, i.e., $\iTheta=\{\tb ,\tm\},$
where $\tb$ and $\tm$ correspond to benign and malicious senders, respectively.
The types $\tb$ and $\tm$ describe the situations where there does not and does exist an adversary, respectively.
The true type $\theta$ is known to the sender, but unknown to the receiver.
Let $\bar{s}^{\rm s}:=(\bss_k)_{k\in\Zp}$ and $\bar{s}^{\rm r}:=(\bsr_k)_{k\in\Zp}$
denote the sender's and receiver's pure \emph{strategy}, respectively.
It is assumed that the receiver's available information about the sender type is only the state, i.e., she cannot observe her instantaneous utility, defined below, nor the sender's action.
Similarly, it is assumed that the sender can observe only the state and her action.
The strategies at the $k$th step with the available information are given by
$\bss_k:\iTheta\times\mc{H}^{\rm s}_k\to\mc{A},$ and $\bsr_k:\mc{H}^{\rm r}_k\to\mc{R}$
where $h^{\rm s}_k\in\mc{H}^{\rm s}_k$ and $h^{\rm r}_k\in\mc{H}^{\rm r}_k$ are histories at the $k$th step given by
$h^{\rm s}_k=(x_{0:k},a_{0:k-1})$ and $h^{\rm r}_k=(x_{0:k},r_{0:k-1}).$
Note that the resulting state trajectory is not Markov since the strategies depend on the entire history.
Because we consider pure strategies, it suffices to consider the state-history dependent strategies
$\ssk:\iTheta\times\mc{X}^{k+1}\to\mc{A}$ and $\srk:\mc{X}^{k+1}\to\mc{R}$,
recursively defined by
\begin{equation}\label{eq:dummy}
\begin{array}{l}
 \ssk(\theta,x_{0:k}):=\bss_k(\theta,x_{0:k},s^{\rm s}_{0:k-1}(x_{0:k-1})),\\
 \srk(x_{0:k}):=\bsr_k(x_{0:k},s^{\rm r}_{0:k-1}(x_{0:k-1})).
\end{array} 
\end{equation}
The strategy profile is denoted by $s:=(s^{\rm s},s^{\rm r}).$
The sender's and receiver's admissible strategy sets are denoted by $\mcss$ and $\mcsr$, respectively.
The set of admissible strategy profiles is denoted by $\mc{S}:=\mcss\times\mcsr$.
Note that, although we do not specify $\mc{S}$ here, it can be taken to be any set of state-history dependent strategies.
While we consider a general strategy set in Sec.~\ref{sec:analysis}, we impose a constraint on $\mc{S}$ in Sec.~\ref{sec:app}.

Once a strategy profile is fixed, the stochastic property of the system is induced.
Construct the canonical measurable space $(\Omega,\mc{F})$ of the MDP with the sender type where $\Omega:=\iTheta \times \Pi_{k=0}^{\infty} (\mc{X}\times\mc{A}\times\mc{R})$ and $\mc{F}$ is its product $\sigma$-algebra~\cite[Chapter~2]{Puterman1994Markov}.
We denote
$\omega=(\theta,(x_0,a_0,r_0),(x_1,a_1,r_1),\ldots)\in\Omega$.
The random variables $\Theta,X_k,A_k,$ and $R_k$ are defined on the measurable space $(\Omega,\mc{F})$ by the projections of $\omega$ such that $\Theta(\omega):=\theta,X_k(\omega):=x_k,$ $A_k(\omega):=a_k,$ $R_k(\omega):=r_k.$
The probability measure on $(\Omega,\mc{F})$, induced by $s$, is denoted by $\Prob^s$,
which satisfies
\begin{equation}\label{eq:dummy}
\left\{
\begin{array}{l}
 \Prob^s(X_0=x_0)=P_0(x_0),\\
 \Prob^s(A_k=a_k|\Theta=\theta,X_{0:k}=x_{0:k})=\delta(a_k,\ssen_k(\theta,x_{0:k})),\\
 \Prob^s(R_k=r_k|X_{0:k}=x_{0:k})=\delta(r_k,\srec_k(x_{0:k})),\\
 \Prob^s(X_{k+1}=x_{k+1}|X_{0:k}=x_{0:k},A_k=a_k,R_k=r_k)\\
 \quad =P(x_{k+1}|x_k,a_k,r_k),\\
 \Prob^s(\Theta=\theta)=\pi_0(\theta)
\end{array}
\right.
\end{equation}
for any $k\in\Zp$ with the initial distribution of the sender type $\pi_0:\iTheta\to[0,1]$.
We denote the conditional probability $\Prob^s(\cdot|\Theta=\theta)$ by $\Prob^s_{\theta}$.
To simplify the notation, we denote the conditional probability mass function with type $\theta$ by
\begin{equation}\label{eq:dummy}
 p^s_\theta(x_{k+1}|x_{0:k}):=\Prob^s_\theta(X_{k+1}=x_{k+1}|X_0=x_0,\ldots,X_k=x_k).
\end{equation}
The expectation with respect to $\Prob^s$ is denoted by $\Exp^s$.




We introduce each player's \emph{belief} on the uncertain variables next.
The receiver's belief at the $k$th step is given by
\begin{equation}\label{eq:dummy}
\begin{array}{l}
 \pir_k(\theta,a_{0:k-1}|x_{0:k},r_{0:k-1})\\
 :=\Prob^s(\Theta=\theta,A_{0:k-1}=a_{0:k-1}| X_{0:k}=x_{0:k},R_{0:k-1}=r_{0:k-1})
\end{array}
\end{equation}
for $k\in\Zp$.
The belief can be recursively computed by Bayes' rule
\begin{equation}\label{eq:dummy}
 \begin{array}{l}
 \pir_{k+1}(\theta,a_{0:k}|x_{0:k+1},r_{0:k}) =\delta(a_k,\ssk(\theta,x_{0:k}))\\
 \times\dfrac{
 P(x_{k+1}|x_k,\ssk(\theta,x_{0:k}),r_k)\pir_k(\theta,a_{0:k-1}|x_{0:k},r_{0:k-1})
 }
 {
 \sum_{\phi\in\iTheta}P(x_{k+1}|x_k,\ssk(\phi,x_{0:k}),r_k)\pir_k(\phi,a_{0:k-1}|x_{0:k},r_{0:k-1})
 }
\end{array}
\end{equation}
when the denominator is nonzero.
To simplify notation, we introduce the receiver's belief only of the sender type:
\begin{equation}\label{eq:dummy}
 \pir_k(\theta|x_{0:k}):=\pir_k(\theta,s^{\rm s}_{0:k-1}(\theta,x_{0:k-1})|x_{0:k}, s^{\rm r}_{0:k-1}(x_{0:k-1})),
\end{equation}
which follows Bayes' rule
\begin{equation}\label{eq:Bayes}
\begin{array}{l}
 \pir_{k+1}(\theta|x_{0:k+1})\\
 =
 \dfrac{
 P(x_{k+1}|x_k,\ssk(\theta,x_{0:k}),\srk(x_{0:k}))\pir_k(\theta|x_{0:k})
 }
 {
 \sum_{\phi\in\iTheta}P(x_{k+1}|x_k,\ssk(\phi,x_{0:k}),\srk(x_{0:k}))\pir_k(\phi|x_{0:k})
 }.
\end{array}
\end{equation}
The sender's belief can similarly be defined and is denoted by $\pis_k(r_{0:k-1}|\theta,x_{0:k},a_{0:k-1})$.

In Sec.~\ref{sec:analysis}, the initial beliefs are assumed to be known to both players, i.e., we make the common prior assumption.
Since we consider pure strategies, $r_{0:k-1}$ is uniquely determined by $x_{0:k-1}$ once the strategy is fixed.
Hence, the sender's belief does not appear explicitly in Sec.~\ref{sec:analysis}.
On the other hand, in Sec.~\ref{sec:app}, we consider the case where the initial belief is unknown to the sender, modeling the possibility of \emph{bluffing}.
Let $U^{\rm s}:\iTheta\times\mc{X}\times\mc{A}\times\mc{R}\to\mathbb{R}$ be the sender's instantaneous utility.
For a given strategy profile $s\in\mc{S}$ and type $\theta\in\iTheta$, the sender's expected average utility at the $k$th step with the horizon length $T$ is given by
\begin{equation}
 \begin{array}{l}
 \bar{U}^{\rm s}_{k,T}(s_{k:k+T}|\theta,x_{0:k})\\
 \displaystyle{
  := \Exp^s\left[\dfrac{1}{T+1} \left. \sum_{\tau=k}^{k+T} U^{\rm s}(\Theta,X_\tau,s^{\rm s}_{\tau}(\Theta,X_{0:\tau}),s^{\rm r}_\tau(X_{0:\tau}))\right|\theta,x_{0:k} \right].
 }
 \end{array}
\end{equation}
Similarly, with the receiver's instantaneous utility given by $U^{\rm r}:\iTheta\times\mc{X}\times\mc{A}\times\mc{R}\to\mathbb{R}$,
the receiver's expected average utility at the $k$th step with the horizon length $T$ is given by
\begin{equation}
\begin{array}{l}
 \bar{U}^{{\rm r}}_{k,T}(s_{k:k+T}|x_{0:k})\\
 \displaystyle{
  := \Exp^s\left[\dfrac{1}{T+1} \left. \sum_{\tau=k}^{k+T} U^{\rm r}(\Theta,X_\tau,s^{\rm s}_{\tau}(\Theta,X_{0:\tau}),s^{\rm r}_\tau(X_{0:\tau}))\right|x_{0:k} \right].
 }
\end{array}
\end{equation}
We denote the limits by
$(\bar{U}^{{\rm s}}_k,\bar{U}^{{\rm r}}_k):=\lim_{T\to\infty} (\bar{U}^{\rm s}_{k,T},\bar{U}^{{\rm r}}_{k,T})$
assuming they exist.
Under this notation, the strategy profile $s=(\ssen,\srec)$ is said to be a \emph{perfect Bayesian equilibrium} (PBE) if
\begin{equation}
\left\{
\begin{array}{l}
\ssen_{k:\infty} \in \BRs_k(\srec_{r:\infty}|\theta,x_{0:k}), \quad \forall \theta\in\iTheta,\\
\srec_{k:\infty} \in \BRr_k(\ssen_{k:\infty}|x_{0:k})
\end{array}
\right.
\end{equation}
for any $k\in \Zp$ and $x_{0:k}\in\mc{X}^{k+1}$
where $\BRs_k$ and $\BRr_k$ are best responses defined by
\begin{equation}\label{eq:dummy}
 \begin{array}{l}
  \BRs_k(\srec_{k:\infty}|\theta,x_{0:k}) 
 := \displaystyle{ \argmax_{\tssen_{k:\infty} \in \mcss_{k:\infty}}\ \bar{U}^{{\rm s}}_k((\tssen_{k:\infty},\srec_{k:\infty})|\theta,x_{0:k})},\\
  \BRr_k(\ssen_{k:\infty}|x_{0:k}) := \displaystyle{\argmax_{\tsrec_{k:\infty} \in \mcsr_{k:\infty}}\ \bar{U}^{{\rm r}}_k((\ssen_{k:\infty},\tsrec_{k:\infty})|x_{0:k})}.
 \end{array}
\end{equation}
Note that, our analysis can be extended to the case of general objective functions rather than expected average utilities as long as the adversary with the utilities avoids being detected, which is formally stated in Definition~\ref{def:det_sens_util} below.

We define the game formulated above by
\begin{equation}\label{eq:game1}
 \mc{G}_1:=(\mc{M},\mc{S},U,\iTheta,\pi_0),
\end{equation}
where the initial belief is common information.
This game belongs to the class of incomplete, imperfect, and asymmetric information stochastic games.
Owing to the existence of the type $\theta,$ which is unknown to the receiver, the information is incomplete.
Because the actions taken by each player are unobservable to the opponent, the information is imperfect and asymmetric.
Although investigating existence and computing equilibria of the game are challenging,
we discuss properties of equilibria on the premise that they exist and are given because our interest here lies in the consequences for the threat.

\section{Analysis: Asymptotic Security}
\label{sec:analysis}

In this section, we analyze asymptotic behaviors of beliefs and actions when the adversary avoids being detected.
It is shown that the system is guaranteed to be secure in an asymptotic manner as long as the defender possesses an effective counteraction.

\subsection{Belief's Asymptotic Behavior}
\label{subsec:conv}

The random variable of the belief on the type $\theta\in\iTheta$ at the $k$th step $\pi^{\theta}_k:\Omega\to[0,1]$ is given by
\begin{equation}\label{eq:dummy}
 \pi^{\theta}_k(\omega):=\pir_k(\theta|X_{0:k}(\omega)).
\end{equation}
Recall that $\pi^{\theta}_k$ represents the defender's confidence on existence of an attacker.
If the belief is low in spite of existence of malicious signals, this means that the Bayesian defense mechanism is deceived.
Because we are interested in whether the Bayesian defense mechanism is permanently deceived, or not, we examine asymptotic behavior of the belief.

We first investigate increment of the belief sequence.
The following lemma is key to our analysis.
\begin{lem}\label{lem:submar}
Consider the game $\mc{G}_1$.
The belief of the true type $\pi^\theta_k$ is a submartingale with respect to the probability $\Prob^s_{\theta}$ and the filtration $\sigma(X_{0:k})$ for any type $\theta$ and strategy profile $s$.
\end{lem}


Lemma~\ref{lem:submar} roughly implies that the expectation of the belief on the true type is non-decreasing.
As a direct conclusion of this lemma, the following theorem holds.
\begin{theorem}\label{thm:conv}
Consider the game $\mc{G}_1$.
There exists an integrable random variable $\pi^\theta_\infty:\Omega\to[0,1]$ such that
\begin{equation}\label{eq:dummy}
 \lim_{k\to\infty}\pi^\theta_k = \pi^\theta_\infty\quad \Prob^s_\theta{\rm -a.s.}
\end{equation}
for any type $\theta$ and strategy profile $s$.
\end{theorem}

\begin{figure}[t]
\centering
\includegraphics[width = .98\linewidth]{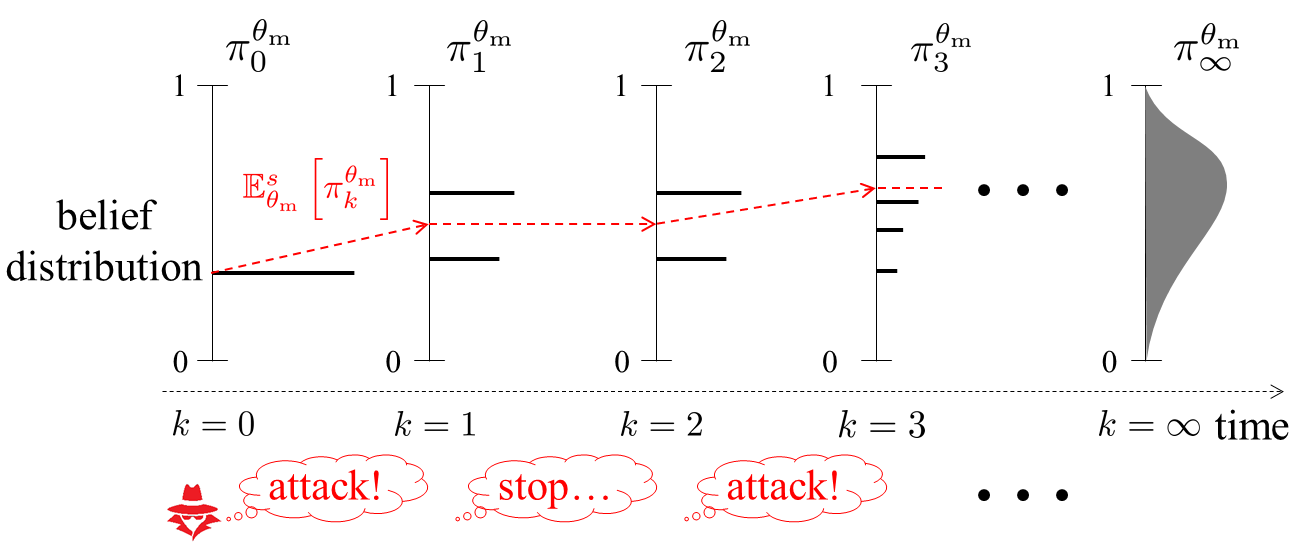}
\caption{
Distributions of the belief sequence when there exists an attacker.
Lemma~\ref{lem:submar} and Theorem~\ref{thm:conv} claim that its expectation is non-decreasing over time and the belief has a limit.
When the adversary stops the attack, the belief is invariant.
}
\label{fig:belief_conv}
\end{figure}

Theorem~\ref{thm:conv} implies that the belief has a limit even if an intermittent attack is executed.
Fig.~\ref{fig:belief_conv} depicts the distributions of the belief sequence when there exists an attacker.
Owing to the model knowledge, if the adversary stops the attack at some time step then the belief is invariant, which is illustrated as the transition of the belief at $k=1$ in Fig.~\ref{fig:belief_conv}.
Moreover, the expectation of the belief is non-decreasing over time as claimed by Lemma~\ref{lem:submar}.
Thus, there exists a limit $\pi^{\tm}_{\infty}$ as shown at the right of Fig.~\ref{fig:belief_conv}.

We next investigate the limit.
An undesirable limit is $\pi^\theta_\infty=0$, which means that \emph{the defender is completely deceived.}
We show that \emph{this does not happen as long as the initial belief is nonzero.}
The following lemma holds.
\begin{lem}\label{lem:log_submar}
Consider the game $\mc{G}_1$.
If $\pi^\theta_0>0$ for any type $\theta,$ then $\log(\pi^\theta_k)$ with any basis converges $\Prob^s_\theta$-almost surely to an integrable random variable as $k\to\infty$ for any type $\theta$ and strategy profile $s$.
\end{lem}

Lemma~\ref{lem:log_submar} leads to the following theorem.
\begin{theorem}\label{thm:b_zero}
Consider the game $\mc{G}_1$.
If $\pi^\theta_0>0$ then
\begin{equation}\label{eq:dummy}
 \pi^\theta_\infty>0\quad \Prob^s_\theta{\rm -a.s.}
\end{equation}
for any type $\theta$ and strategy profile $s$.
\end{theorem}
Theorem~\ref{thm:b_zero} implies that the complete deception described by $\pi^\theta_\infty=0$ does not occur.

{\it Remark:}
Theorems~\ref{thm:conv} and~\ref{thm:b_zero} can heuristically be justified from an information-theoretic perspective as follows.
Suppose that the state sequence $x_{0:k}$ is observed at the $k$th step.
Then the belief is given by
\begin{equation}\label{eq:pi_KL}
 \begin{array}{cl}
 \pi_k(\theta|x_{0:k})
 \hs = \dfrac{\pi_0(\theta)}
 {\sum_{\phi\neq \theta} \frac{p^s_\phi(x_{0:k})}{p^s_\theta(x_{0:k})} \pi_0(\phi) + \pi_0(\theta) }\\
 \hs = \dfrac{\pi_0(\theta)}
 {
 \sum_{\phi\neq \theta} {\rm exp}(kS^\phi_k(x_{0:k})) \pi_0(\phi) + \pi_0(\theta)
 }
 \end{array}
\end{equation}
where
$p^s_{\theta}(x_{0:k})$ and $p^s_{\phi}(x_{0:k})$ are the joint probability mass functions of $x_{0:k}$ with respect to $\Prob^s_\theta$ and $\Prob^s_\phi$, respectively, and
\begin{equation}\label{eq:dummy}
 S^\phi_k(x_{0:k}):=\dfrac{1}{k} \sum_{i=1}^k \log \dfrac{p^s_{\phi}(x_i|x_{0:i-1}) }
 {p^s_{\theta}(x_i|x_{0:i-1})}.
\end{equation}
Assuming that
$p^s_\theta(x_k|x_{0:k-1})$
approaches a stationary distribution
$p^s_\theta(x)$ on $\mc{X}$
and SLLN can be applied,
we have
\begin{equation}\label{eq:dummy}
\lim_{k\to\infty} S^\phi_k=\mathbb{E}_{x\sim p^s_\theta}
 \left[
 \log p^s_{\phi}(x)/p^s_\theta(x)
 \right]
 = -D_{\rm KL}(p^s_\theta||p^s_\phi)
\end{equation}
where $D_{\rm KL}$ denotes the Kullback-Leibler divergence.
Since $D_{\rm KL}$ is nonnegative for any pair of distributions, $S^\phi_k$ converges to a nonpositive number, which results in convergence of $\pi^\theta_k$.
If $p^s_\theta\neq p^s_\phi$ for any $\phi\in\iTheta\setminus\{\theta\}$, the limit of $S^\phi_k$ becomes negative, and hence $\lim_{k\to\infty}\exp(k S^\phi_k(x_{0:k}))=0,$ which leads to
\begin{equation}\label{eq:dummy}
 \begin{array}{cl}
 \displaystyle{\lim_{k\to \infty} \pi_k(\theta|x_{0:k})}
 \hs = \dfrac{\pi_0(\theta)}
 {
 \displaystyle{
 \sum_{\phi\neq \theta} \lim_{k\to\infty} {\rm exp}(kS^\phi_k(x_{0:k})) \pi_0(\phi) + \pi_0(\theta)
 }
 }\\
 \hs = 1.
 \end{array}
\end{equation}
Thus, the belief of the true type converges to one.
Such a convergence property of the Bayesian estimator on the true parameter, referred to as Bayesian consistency, has been investigated mainly in the context of statistics~\cite{Diaconis1986On,Walker2004New}.
In this sense, Theorems~\ref{thm:conv} and~\ref{thm:b_zero} can be regarded as another representation of Bayesian consistency.
However, note again that this discussion is not a rigorous proof but a heuristic justification because the state is essentially non-i.i.d. and even non-ergodic in our game-theoretic formulation.

\subsection{Asymptotic Security}

It has turned out that the belief has a positive limit.
To clarify our interest, we define the notion of detection-averse utilities.
\begin{defin}\label{def:det_sens_util}
{\bf (Detection-averse Utilities)}
A pair $(\Us,\Ur)$ in the game $\mc{G}_1$ are detection-averse utilities when
\begin{equation}\label{eq:det_sens_str}
 \pi^{\tm}_\infty < 1\quad \Prob^s_{\tm}{\rm -a.s.}
\end{equation}
for any PBE $s$.
\end{defin}
Definition~\ref{def:det_sens_util} characterizes utilities where the malicious sender avoids having the defender form a firm belief on the existence of an attacker.
An example of detection-averse utilities is given in Appendix~\ref{app:ex_detection-averse}.
Naturally, strategies reasonable for the attacker should be detection-averse as long as the defender possesses an effective counteraction.
If the utilities of interest are not detection-averse, this means that the defense mechanism cannot cope with the attack because the attacker is not afraid to reveal herself.
For protecting such systems, appropriate counteractions should be implemented beforehand.

Suppose that there is an effective countermeasure, and hence the utilities are detection-averse.
A simple malicious sender's strategy that satisfies~\eqref{eq:det_sens_str} is to imitate the benign sender's strategy after a finite number of time steps.
We give a formal definition of such strategies.
\begin{defin}\label{def:asymp_benign}
{\bf (Asymptotically Benign Strategy)}
A strategy profile $s$ in the game $\mc{G}_1$ is asymptotically benign when
\begin{equation}\label{eq:dummy}
 \lim_{k\to\infty}\delta\left(A^{\tm}_k,A^{\tb}_k\right) = 1 \quad \Prob^s_{\tm}{\rm -a.s.}
\end{equation}
where $A^{\theta}_k$ is the action taken by the sender with the type $\theta$ defined by
$A^{\theta}_k:=\ssen_k(\theta,X_{0:k}).$
\end{defin}


The objective of this subsection is to show that Bayesian defense mechanisms can restrict all reasonable strategies to be asymptotically benign as long as an effective countermeasure is implemented.

As a preparation for proving our main claim, we investigate the asymptotic behavior of state transition.
From Theorems~\ref{thm:conv} and~\ref{thm:b_zero}, we can expect that the state eventually loses information on the type, which is justified by the following lemma.
\begin{lem}\label{lem:pp}
Consider the game $\mc{G}_1$ with detection-averse utilities.
If $\pi^{\tm}_0>0,$ then
\begin{equation}\label{eq:dummy}
 \lim_{k\to\infty}\left|p^s_{\tm}(X_{k+1}|X_{0:k})-p^s_{\tb}(X_{k+1}|X_{0:k})\right| = 0\quad \Prob^s_{\tm}{\rm -a.s.}
\end{equation}
for any PBE $s$.
\end{lem}

Under Assumption~\ref{assum:input_obs}, which eliminates the possibility of stealthy attacks, Lemma~\ref{lem:pp} implies that the actions themselves must be identical.
This fact yields the following theorem, one of the main results in this paper.
\begin{theorem}\label{thm:main_asymp_sec}
Consider the game $\mc{G}_1$ with detection-averse utilities.
Let Assumption~\ref{assum:input_obs} hold and assume $\pi^{\tm}_0>0$.
Then, every PBE of $\mc{G}_1$ is asymptotically benign.
\end{theorem}


Theorem~\ref{thm:main_asymp_sec} implies that the malicious sender's action converges to the benign action.
Equivalently, an attacker necessarily behaves as a benign sender after a finite number time steps.
Therefore, the system is guaranteed to be secure in an asymptotic manner, i.e., Bayesian defense mechanisms \emph{can} prevent deception in an asymptotic sense.
This result indicates the powerful defense capability achieved by model knowledge.



\section{Application: Analysis of Defensive Deception utilizing Asymmetric Recognition}
\label{sec:app}

\subsection{Idea of Defensive Deception using Bluffing}

The result in Section~\ref{sec:analysis} claims that the defender, namely, the Bayesian defense mechanism, always wins in an asymptotic manner when the stochastic model of the system is available and \emph{the vulnerability to be exploited for intrusion is known and modeled.}
The latter condition is quantitatively described by the condition $\pi_0(\tm)>0$.
Although the derived result proves a quite powerful defense capability, it is also true that it is almost impossible to be aware of all possible vulnerabilities in advance.
Moreover, it is also challenging to implement effective countermeasures for all scenarios and to compute the equilibrium of the dynamic game.

In this section, as an application of the finding in the previous section, we consider \emph{defensive deception} using bluffing that utilizes asymmetric recognitions between the attacker and the defender.
Suppose that an attacker exploits a vulnerability of which the defender is unaware but the attacker is unaware of the defender's unawareness.
Then their recognition becomes asymmetric in the sense that the attacker does not correctly recognize the defender's recognition of the vulnerability.
This situation naturally arises in practice because the defender's recognition is private information.
By utilizing the asymmetric recognition, the defender can possibly deceive the attacker such that the attacker believes that the defender might be aware of the vulnerability and carrying out effective counteractions.
Specifically, we consider the bluffing strategies where the system's state does not possess information about the defender's belief.
For instance, if the defender chooses the reactions that affect only the players' utilities without influence to the system, the state is independent of the reaction.
By concealing the defender's unawareness, the defender's recognition, which is quantified by her belief, is completely unknown to the attacker over time.

The defensive deception is possibly able to force the attacker to withdraw even if the defender is actually unaware of the exploited vulnerability.
For instance, consider the example in Sec.~\ref{subsec:sketch} and suppose that emergency shutdown of the system can be carried out by the defender.
Suppose also that the attacker wants to keep administrative privileges of the PLC.
In this case, the attacker may rationally terminate her evasive maneuvers after a finite number of time steps due to the risk of sudden shutdown.
The objective of this section is to show that the hypothesis is true in a formal manner.

\subsection{Reformulation using Type Structure}

\begin{figure}[t]
\centering
\includegraphics[width=.95\linewidth]{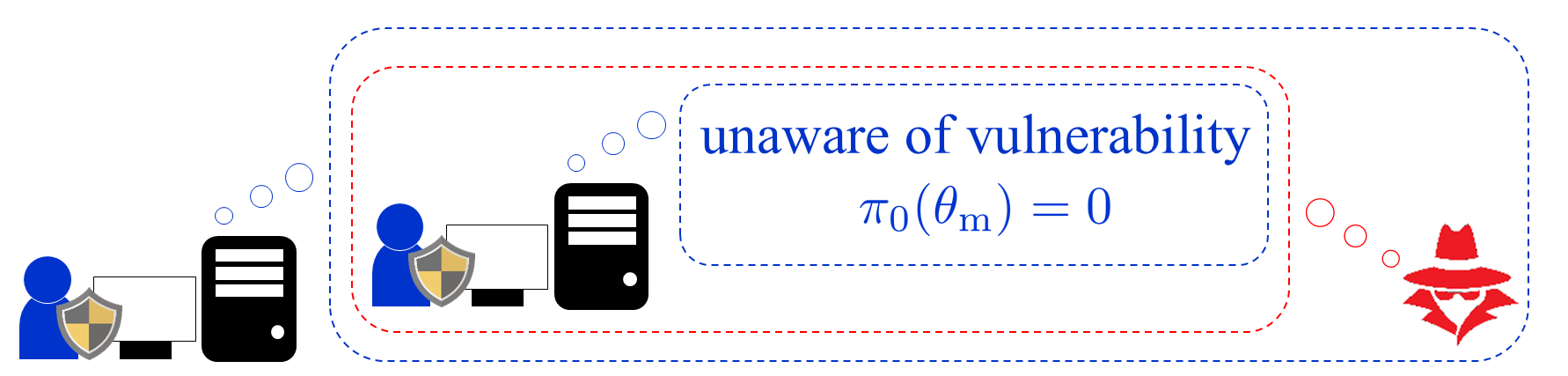}
\caption{
The attacker's belief of the defender's belief with symmetric recognition, which is the case of the game $\mc{G}_1$.
The attacker is aware of the fact that the defender is unaware of the vulnerability.
Moreover, the defender is aware of the attacker's awareness.
}
\label{fig:sym_rec}
\end{figure}

\begin{figure}[t]
\centering
\includegraphics[width=.95\linewidth]{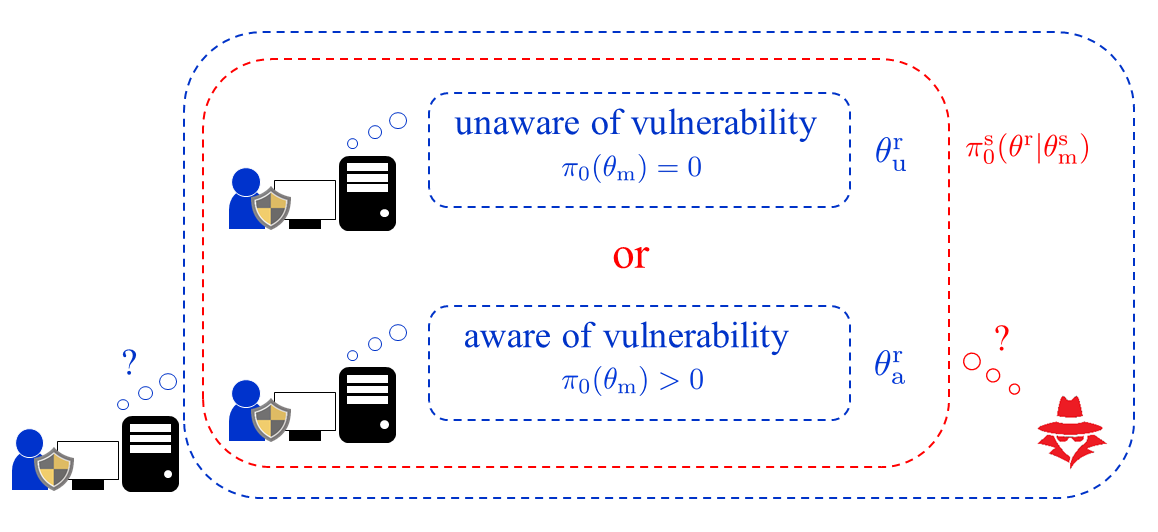}
\caption{
The attacker's belief of the defender's belief with asymmetric recognition.
Because the defender's true belief is unknown to the attacker, the attacker forms a belief on both cases that the defender is aware or unaware of the vulnerability.
Moreover, the defender forms a belief on the attacker's belief.
This process induces the notion of belief hierarchy.
}
\label{fig:asym_rec}
\end{figure}

The situation of interest in this section is that the defender is unaware of the vulnerability to be exploited but the attacker is \emph{not} necessarily aware of this unawareness.
To address the uncertainty on defender's recognition, the attacker forms her belief on the defender's belief.
Fig.~\ref{fig:sym_rec} illustrates the attacker's belief on the defender's belief with the common prior assumption, i.e., the initial defender's belief is known to the attacker, which has been made in the previous section.
In this case, the attacker has a firm belief that the defender is unaware of the vulnerability.
On the other hand, Fig.~\ref{fig:asym_rec} illustrates the attacker's belief without the common prior assumption.
Then the attacker's belief is no longer firm as depicted by the figure.
In addition, because of the lack of the common prior assumption, the defender also forms another belief on the attacker's belief on the defender's belief on the existence of an attacker.
This procedure repeats indefinitely and induces infinitely many beliefs.

The notion of \emph{belief hierarchy} has been proposed to handle the infinitely many beliefs~\cite{Mertens1985Formulation,Dekel2015Epistemic,Zamir2009Bayesian}.
A belief hierarchy is formed as follows.
Let $\Delta(\cdot)$ denote the set of probability measures over a set.
The first-order initial belief is given as $\pi^1_0 \in \Delta(\iTheta)$, which describes the defender's initial belief on existence of the attacker.
The second-order initial belief is given as $\pi^2_0\in \Delta(\Delta(\iTheta))$, which describes the attacker's initial belief on the defender's first-order belief.
In a similar manner, the belief at any level is given, and the tuple of beliefs at all levels is referred to as a belief hierarchy.

To handle belief hierarchies, the \emph{Mertens-Zamir model} has been introduced~\cite{Mertens1985Formulation,Dekel2015Epistemic,Zamir2009Bayesian}.
The model considers \emph{type structure}, in which a belief hierarchy is embedded.
A type structure consists of players, sets of types, and initial beliefs.
In particular, a type structure for our situation of interest can be given by
\begin{equation}\label{eq:type_str}
 \mc{T}=(({\rm s},{\rm r}),(\iThetas,\iThetar),(\pis_0,\pir_0))
\end{equation}
where $({\rm s},{\rm r})$ represents the sender and the receiver,
$\iThetas$ and $\iThetar$ represent the sets of player types,
and $\pis_0:\iThetar\times\iThetas\to[0,1]$ and $\pir_0:\iThetas\times\iThetar\to[0,1]$ represent the initial beliefs.
The value $\pis_0(\tr|\ts)$ denotes the sender's initial belief of the receiver type $\tr$ when the sender type is $\ts$, and $\pir_0(\ts|\tr)$ denotes the corresponding receiver's initial belief.
The first-order initial belief is given by $\pi^1_0(\ts)=\pir_0(\ts|\tr)$ for the true receiver type $\tr\in\iThetar$, and the second-order initial belief is given by $\pi^2_0(\pir(\cdot|\tr)|\ts)=\pis_0(\tr|\ts)$ for the true sender type $\ts\in\iThetas$.
By repeating it, the belief at any level of the belief hierarchy can be derived from the type structure.
Importantly, for any reasonable belief hierarchy there exists a type structure that can generate the belief hierarchy of interest.
For a formal discussion, see~\cite{Mertens1985Formulation,Dekel2015Epistemic,Zamir2009Bayesian}.

We model the situation of interest by using the binary type sets:
\begin{equation}\label{eq:dummy}
 \iThetas = \{\tsb,\tsm\},\quad\iThetar = \{\tru,\tra\}.
\end{equation}
While $\tsb$ and $\tsm$ represent benign and malicious senders, respectively,
$\tru$ and $\tra$ represent receivers being unaware and aware of the vulnerability, respectively.
The receiver's initial beliefs are set to
\begin{equation}\label{eq:dummy}
 \pir_0(\tsb|\tru)=1,\quad \pir_0(\tsm|\tru)=0
\end{equation}
and
\begin{equation}\label{eq:dummy}
 \pir_0(\tsb|\tra)=\alpha,\quad \pir_0(\tsm|\tra)=1-\alpha
\end{equation}
with $\alpha\in[0,1)$.
The initial beliefs mean that, the receiver $\tru$ is unaware of the vulnerability and firmly believes that the system is normally operated, while the receiver $\tra$ is aware of the vulnerability and suspects existence of an attacker with probability $1-\alpha$.
The sender's initial beliefs are assumed to be given by
\begin{equation}\label{eq:dummy}
 \pis_0(\tru|\tsb)=1,\quad \pis_0(\tra|\tsb)=0,
\end{equation}
and
\begin{equation}\label{eq:dummy}
 \pis_0(\tru|\tsm)=\beta,\quad \pis_0(\tra|\tsm)=1-\beta
\end{equation}
with $\beta\in[0,1]$.
The malicious sender does not know the true receiver type, i.e., whether the sender is aware of the vulnerability or not.
The given initial beliefs are summarized in Table~\ref{table:initial_belief}.

\begin{table}[t]
\centering
\caption{Initial Beliefs on Opponent Type}
\begin{tabular}{c|cc}
  & $\tsb$ & $\tsm$\\ \hline
 $\pir_0(\cdot|\tru)$ & 1 & 0 \\
 $\pir_0(\cdot|\tra)$ & $\alpha$ & $1-\alpha$
\end{tabular}
\hspace{3mm}
\begin{tabular}{c|cc}
  & $\tru$ & $\tra$\\ \hline
 $\pis_0(\cdot|\tsb)$ & 1 & 0 \\
 $\pis_0(\cdot|\tsm)$ & $\beta$ & $1-\beta$
\end{tabular}
\label{table:initial_belief}
\end{table}

In accordance with the introduction of the type structure, the definition of strategies and the solution concept are needed to be slightly modified.
The contrasting ingredients of the game with symmetric recognition and the one with asymmetric recognition are listed in Table~\ref{table:ingredients}, where those with asymmetric recognition can analogically be defined.
The conditional probability $\Prob^s(\cdot|\Ts=\ts,\Tr=\tr)$, which is the probability measure induced by $\ssen(\ts,\cdot)$ and $\srec(\tr,\cdot)$, is denoted by $\Prob^s_{\ts,\tr}$.
The sender's expected average utility at the $k$th step with the horizon length $T$ is given by
\begin{equation}\label{eq:UsT_G2}
\begin{array}{l}
\bar{U}^{\rm s}_{k,T}(s_{k:k+T}|\ts,x_{0:k}):= \dfrac{1}{T+1}\\
\displaystyle{\small{
  \times\Exp^s\left[\left. \sum_{\tau=k}^{k+T} U^{\rm s}(\Ts,X_\tau,\ssen_{\tau}(\Ts,X_{0:\tau}),\srec_{\tau}(\Tr,X_{0:\tau})) \right| \ts,x_{0:k}\right].
  }
}
\end{array}
\end{equation}
The receiver's expected average utility at the $k$th step with the horizon length $T$ is given by
\begin{equation}\label{eq:dummy}
\begin{array}{l}
\bar{U}^{\rm r}_{k,T}(s_{k:k+T}|\tr,x_{0:k}):=\dfrac{1}{T+1} \\
\displaystyle{\small{
  \times\Exp^s\left[\left. \sum_{\tau=k}^{k+T} U^{\rm r}(\Ts,X_\tau,\ssen_{\tau}(\Ts,X_{0:\tau}),\srec_{\tau}(\Tr,X_{0:\tau})) \right| \tr,x_{0:k}\right].
  }
}
\end{array}
\end{equation}
A strategy $s$ is said to be a PBE when the limit of the utilities $(\bar{U}^{\rm s}_k,\bar{U}^{\rm r}_k)$ satisfies
\begin{equation}\label{eq:dummy}
\left\{
\begin{array}{l}
\ssen_{r:\infty} \in \BRs_k(\srec_{r:\infty}|\ts,x_{0:k}),\quad \forall \ts\in\iThetas,\\
\srec_{k:\infty} \in \BRr_k(\ssen_{k:\infty}|\tr,x_{0:k}),\quad \forall \tr\in\iThetar,
\end{array}
\right.
\end{equation}
for any $k\in\Zp$ and $x_{0:k}\in\mc{X}^{k+1}$
where
\begin{equation}\label{eq:dummy}
\begin{array}{l}
 \BRs_k(\srec_{k:\infty}|\ts,x_{0:k})
 := \displaystyle{ \argmax_{\tssen_{k:\infty} \in \mcss_{k:\infty}}\ \bar{U}^{\rm s}_k((\tssen_{k:\infty},\srec_{k:\infty})|\ts,x_{0:k})},\\
 \BRr_k(\ssen_{k:\infty}|\tr,x_{0:k})
 := \displaystyle{ \argmax_{\tsrec_{k:\infty} \in \mcsr_{k:\infty}}\ \bar{U}^{\rm r}_k((\ssen_{k:\infty},\tsrec_{k:\infty})|\tr,x_{0:k})}.
\end{array}
\end{equation}

\begin{table}[t]
\centering
\caption{Contrasting ingredients of the games with symmetric and asymmetric recognitions}
\begin{tabular}{|c|c|c|} \hline
  & symmetric recognition & asymmetric recognition \\ \hline
 receiver's strategy & $\srec_k(x_{0:k})$ & $\srec_k(\tr,x_{0:k})$ \\ \hline
 sender's belief & N/A & $\pis(\tr|\ts)$ \\ \hline
 receiver's belief & $\pir(\theta)$ & $\pir(\ts|\tr)$ \\ \hline
 sender's utility & $\bar{U}^{\rm s}(s,\theta)$ & $\bar{U}^{\rm s}(s,\ts)$ \\ \hline
 receiver's utility & $\bar{U}^{\rm r}(s)$ & $\bar{U}^{\rm r}(s,\tr)$ \\ \hline
\end{tabular}
\label{table:ingredients}
\end{table}

We define the game formulated above by
\begin{equation}\label{eq:game2}
 \mc{G}_2:=(\mc{M},\mc{S},U,(\iThetas,\iThetar),(\pis_0,\pir_0)),
\end{equation}
where the defender's initial belief is \emph{not} common information in contrast to $\mc{G}_1$.

In the following discussion, we analyze $\mc{G}_2$ through $\mc{G}_1$.
To clarify their relationship, we describe the game $\mc{G}_1$ using the modified formulation.
Define another game
\begin{equation}\label{eq:dummy}
 \hat{\mc{G}}_2:=(\mc{M},\mc{S},U,(\iThetas,\iThetar),(\hpis_0,\pir_0)),
\end{equation}
where
\begin{equation}\label{eq:dummy}
 \hpis_0(\tra|\tsm)=1.
\end{equation}
The initial belief means that the adversary believes that the defender is aware of the vulnerability.
The situation of $\hat{\mc{G}}_2$ is the same as that of $\mc{G}_1$ if the defender is aware of the vulnerability.
Thus, these games lead to the same consequence when the true types are $\tsm$ and $\tra$.
The following lemma holds.

\begin{lem}\label{lem:G1G2_eq}
Consider the games $\mc{G}_1$ and $\hat{\mc{G}}_2$.
For a strategy profile $\hat{s}_2=(\hssen_2,\hsrec_2)$ in $\hat{\mc{G}}_2$, let $s_1=(\ssen_1,\srec_1)$ be a strategy profile in $\mc{G}_1$ such that
\begin{equation}\label{eq:str_restriction}
 \ssen_1:=\hssen_2,\quad \srec_1:=\hsrec_2|_{\tr=\tra}
\end{equation}
where $\hsrec_2|_{\tr=\tra}$ is the restriction of $\hsrec_2$ with $\tr=\tra$.
Then the probability measures induced by $s_1$ and $\hat{s}_2$ are equal when $\ts=\tsm$ and $\tr=\tra$, i.e.,
\begin{equation}\label{eq:dummy}
 \Prob^{s_1}_{\tsm}=\Prob^{\hat{s}_2}_{\tsm,\tra}.
\end{equation}
Also, if $\hat{s}^{\rm s}_{2,k:\infty} \in \BRs_k(\hat{s}^{\rm r}_{2,k:\infty}|\tsm,x_{0:k})$ then $\ssen_{1,k:\infty}\in \BRs_k(\srec_{1,k:\infty}|\tm,x_{0:k})$.
\end{lem}


We extend the notions of detection-averse utilities and asymptotically benign strategies to $\mc{G}_2$.
Our objective is to investigate the effectiveness of the proposed defensive deception.
It is possible to define detection-averse utilities directly using the game $\mathcal{G}_2$ as utilities where the resulting equilibrium leads the adversary to avoid being detected.
However, this definition immediately means that the defensive deception works well, and any results from the definition cannot show its effectiveness.
Instead, we say that utilities in $\mc{G}_2$ are detection-averse when the adversary avoids being detected if she is certain that the defender is aware of the vulnerability.
\begin{defin}\label{def:detection-averse_G2}
{\bf (Detection-averse Utilities in $\mc{G}_2$)}
A pair of utilities $(U^{\rm s},U^{\rm r})$ in the game $\mc{G}_2$ are detection-averse utilities when
\begin{equation}\label{eq:firm_belief_G2}
 \lim_{k\to\infty} \pir_k(\tsm|\tra)<1\quad \Prob^s_{\tsm,\tra}{\rm -a.s.}
\end{equation}
for any PBE $s$ of $\hat{\mc{G}}_2$.
\end{defin}

Note that Definition~\ref{def:detection-averse_G2} is a necessary requirement to make the game interesting, because the adversary is not afraid of being detected at all without this condition.


Next, we define desirable strategies that should be achieved by Bayesian defense mechanisms.
We say a strategy in $\mc{G}_2$ to be asymptotically benign when it becomes benign regardless of the defender's awareness.
\begin{defin}\label{def:asymp_benign_G2}
{\bf (Asymptotically Benign Strategies in $\mc{G}_2$)}
A strategy profile $s$ in the game $\mc{G}_2$ is asymptotically benign when
\begin{equation}\label{eq:asymp_benign_G2}
 \lim_{k\to\infty}\delta\left(A^{\tm}_k,A^{\tb}_k\right) = 1 \quad \Prob^s_{\tsm,\tr}{\rm -a.s.}
\end{equation}
for any $\tr\in\iThetar$.
\end{defin}

Note that Definition~\ref{def:asymp_benign_G2} requires the strategy to be asymptotically benign for any $\tr\in\iThetar$.
In other words, the strategy is needed to be asymptotically benign even if the defender is unaware of the vulnerability.

\subsection{Passively Bluffing Strategies}

We expect that there exists a chance of preventing attacks that exploit unnoticed vulnerabilities if the state does not possess information about the defender's recognition.
To formally verify this expectation, we define passively bluffing strategies.
\begin{defin}\label{def:passive_bluffing}
{\bf (Passively Bluffing Strategies)}
A strategy profile $s$ in $\mc{G}_2$ is a passively bluffing strategy profile when the sender's belief satisfies
\begin{equation}\label{eq:belief_invariant}
 \pis_k(\tr|X_{0:k},\ts)=\pis_0(\tr|\ts)\quad \Prob^s_{\ts,\tr}{\rm -a.s.}
\end{equation}
for any $\ts\in\iThetas,\tr\in\iThetar,$ and $k\in\Zp$.
A strategy profile set $\mc{S}$ in $\mc{G}_2$ is a passively bluffing strategy set when its all elements are passively bluffing.
\end{defin}

Definition~\ref{def:passive_bluffing} requires the sender's belief to be invariant over time.
If the strategy is passively bluffing, the adversary cannot identify whether the defender is aware of the exploited vulnerability or not even in an asymptotic sense.
Note that the introduced passively bluffing strategies can be regarded as a commitment.
It is well known that restricting feasible strategies, referred to as commitment, can be beneficial in a game~\cite{Heifetz2012Commitment,Letchford2014Value}.
In what follows, we investigate the effectiveness of the specific commitment.

Passively bluffing strategies can relax the condition for asymptotically benign strategies.
The following lemma holds.
\begin{lem}\label{lem:relax_asymp_benign}
Consider the game $\mc{G}_2$.
If a passively bluffing strategy profile $s$ satisfies
\begin{equation}\label{eq:asymp_benign_tra}
 \lim_{k\to\infty}\delta\left(A^{\tm}_k,A^{\tb}_k\right) = 1 \quad \Prob^s_{\tsm,\tra}{\rm -a.s.}
\end{equation}
then $s$ is asymptotically benign.
\end{lem}

The difference between~\eqref{eq:asymp_benign_G2} and~\eqref{eq:asymp_benign_tra} is the required receiver type.
Lemma~\ref{lem:relax_asymp_benign} implies that if a passively bluffing strategy profile is asymptotically benign when the receiver is aware of the vulnerability then the strategy is needed to be asymptotically benign even when the receiver is unaware of the vulnerability.

\emph{Remark:}
Although Definition~\ref{def:passive_bluffing} depends not only on the receiver's strategy but also on the sender's strategy for generality, the bluffing should be realized only by the defender in practice.
A simple defender's approach to achieving the bluffing is to choose reactions that do not influence the system's behavior.
Let $\mc{R}_{\rm pb}\subset \mc{R}$ be the set of reactions such that the system's dynamics is independent of the reaction, i.e., the transition probability satisfies
\begin{equation}\label{eq:sys_indep_r}
 P(x'|x,a,r)=P(x'|x,a,r')
\end{equation}
for any $x'\in\mc{X},x\in\mc{X}$, $a\in\mc{A}$, $r\in\mc{R}_{\rm pb}$, $r'\in\mc{R}_{\rm pb}$.
If the receiver's strategy takes only reactions in $\mc{R}_{\rm pb},$ every strategy profile is passively bluffing.
Indeed, because the transition probability is independent of $r\in\mc{R}_{\rm pb}$, the probability distribution of the state is independent of $\tr$.
Thus, from Bayes' rule, we have
\begin{equation}\label{eq:dummy}
\begin{array}{cl}
 \pis_{k}(\tr|x_{0:k},\ts) \hs = \dfrac{p^s_{\ts,\tr}(x_{0:k}) \pis_0(\tr|\ts)}
 {\sum_{\phi^{\rm r}\in\iThetar} p^s_{\ts,\phi^{\rm r}}(x_{0:k}) \pis_0(\phi^{\rm r}|\ts) }\\
 \hs = \dfrac{p^s_{\ts}(x_{0:k}) \pis_0(\tr|\ts)}
 {\sum_{\phi^{\rm r}\in\iThetar} p^s_{\ts}(x_{0:k}) \pis_0(\phi^{\rm r}|\ts) }\\
 \hs = \dfrac{\pis_0(\tr|\ts)}
 {\sum_{\phi^{\rm r}\in\iThetar} \pis_0(\phi^{\rm r}|\ts) }\\
 \hs = \pis_0(\tr|\ts)
\end{array}
\end{equation}
when $p^s_{\ts,\tr}(x_{0:k})\neq 0$.
An example of such reactions is just analyzing the network log and raising an alarm inside the operation room without applying control on the system itself.
Note that the reaction still affects the players' decision making through their utility functions, even if~\eqref{eq:sys_indep_r} holds.

\subsection{Analysis}

Our expectation can be described in a quantitative form based on the definition of passively bluffing strategies, which lead to a simple representation of the sender's utility.
If $s$ is passively bluffing, the sender's belief is invariant over time.
Hence, the sender's utility with infinite horizon is given by
\begin{equation}\label{eq:bUs_weighted_sum}
\begin{array}{l}
\displaystyle{ 
 \bUs_k(s_{k:\infty}|\tsm,x_{0:k})
 =\sum_{\tr\in\iThetar} \bUs_{k,\tr}(s_{k:\infty}|\tsm,x_{0:k}) \pis_0(\tr|\tsm)
}
\end{array}
\end{equation}
where
\begin{equation}\label{eq:dummy}
 \begin{array}{l}
  \bUs_{k,\tr}(s_{k:\infty}|\tsm,x_{0:k}):=\lim_{T\to\infty} \dfrac{1}{T+1} \\
  \displaystyle{
  \times \Exp^s_{\tr}\left[\left. \sum_{\tau=k}^{k+T} U^{\rm s}(\tsm,X_\tau,\ssen_{\tau}(\tsm,X_{0:\tau}),\srec_{\tau}(\tr,X_{0:\tau})) \right|x_{0:k}\right].
  }
 \end{array}
\end{equation}
Note that $\bUs_{k,\tra}$ and $\bUs_{k,\tru}$ denote the sender's utilities of the two cases where the defender is aware and unaware of the vulnerability, respectively.
Thus~\eqref{eq:bUs_weighted_sum} implies that the sender's utility is simply given as a sum weighted by her initial beliefs when the strategy is passively bluffing.
Therefore, we can expect that the sender possibly stops the execution in the middle of the attack if $\pis_0(\tra|\tsm)$ is sufficiently large.
We show the existence of such sender's initial belief.
Note that $\pis_0(\tra|\tsm)=1$
is the trivial case, and thus we assume that sender's initial beliefs that are strictly less than one.

First, we rephrase the result in Sec.~\ref{sec:analysis}.
Let $\mc{S}_{\rm nab}$ denote the set of non-asymptotically-benign strategies in $\mc{G}_2$.
Our aim here is to show that the set of PBE of $\mc{G}_2$ does not overlap with $\mc{S}_{\rm nab}$ when the attacker strongly believes that the defender is aware of the vulnerability.
It suffices to show that there is no overlap between the set of PBE of $\mc{G}_2$ and $\mc{S}_{\rm nab}^{\ast}:=\mc{S}_{\rm nab}\cap\mc{S}^{\ast}$ where
\begin{equation}\label{eq:dummy}
 \begin{array}{l}
 \mc{S}^{\ast} :=\{ (\ssen,\srec)\in\mc{S}: \ssen_{k:\infty}\in\BRr_k(\srec_{k:\infty}|\tsb,x_{0:k}),\\
  \srec_{k:\infty}\in\BRr_k(\ssen_{k:\infty}|\tr,x_{0:k}),\forall \tr\in\iThetar, k\in\Zp, x_{0:k}\in\mc{X}^{k+1}\},
 \end{array}
\end{equation}
where the benign sender and the receiver with any type take their best response strategies.
Note that, $\mc{S}_{\rm nab}$ and $\mc{S}^{\ast}$ of the games $\mc{G}_2$ and $\hat{\mc{G}}_2$ are identical because the sets are independent of the malicious sender's belief.
The following lemma is another description of the claim of Theorem~\ref{thm:main_asymp_sec} with respect to $\bUs_{\tra}$ and $\mc{S}_{\rm nab}^{\ast}$.
\begin{lem}\label{lem:rephrase}
Consider the game $\hat{\mc{G}}_2$ with detection-averse utilities.
Let Assumption~\ref{assum:input_obs} hold.
For any strategy profile $s=(\ssen,\srec)$ in $\mc{S}_{\rm nab}^{\ast}$, there exists $\tssen \in \mcss$ such that
\begin{equation}\label{eq:non_PBE}
D_{\tra}(s,\tssen)>0
\end{equation}
holds
where
\begin{equation}\label{eq:D_tr}
  D_{\tr}(s,\tssen):=\bUs_{\tr}((\tssen,\srec),\tsm)-\bar{U}^{\rm s}_{\tr}(s,\tsm).
\end{equation}
\end{lem}

Lemma~\ref{lem:rephrase} implies the existence of a function
\begin{equation}\label{eq:func_g}
 g:\mc{S}_{\rm nab}^{\ast}\to\mcss\quad {\rm s.t.}\quad D_{\tra}(s,g(s))>0
\end{equation}
for any $s\in\mc{S}^{\ast}_{\rm nab}.$
Thus we have $\gamma\geq 0$ where
\begin{equation}\label{eq:inf_sup}
 \gamma := \inf_{s\in \mc{S}_{\rm nab}^{\ast}} D_{\tra}(s,g(s)).
\end{equation}
We here make an assumption that $D_{\tra}(s,g(s))$ is uniformly lower bounded by a positive value.
\begin{assum}\label{assum:uniformly_bounded}
For the game $\hat{\mc{G}}_2$, there exists $g$ in~\eqref{eq:func_g} such that the infimum~\eqref{eq:inf_sup} is positive, i.e., $\gamma>0$.
\end{assum}

Assumption~\ref{assum:uniformly_bounded} eliminates the case where the difference between the sender's utilities achievable by asymptotically benign strategies and non-asymptotically-benign strategies is infinitesimally small.

The following theorem, the main result of this section, holds.
\begin{theorem}\label{thm:bluffing}
Consider the game $\mc{G}_2$ with detection-averse utilities and a passively bluffing strategy set.
Let Assumptions~\ref{assum:input_obs} and~\ref{assum:uniformly_bounded} hold.
Then, there exists a sender's initial belief $\pis_0(\tra|\tsm)<1$ such that every PBE of $\mc{G}_2$ is asymptotically benign.
\end{theorem}

Theorem~\ref{thm:bluffing} implies that the system can possibly be protected by passively bluffing strategies if the attacker strongly believes that the defender is aware of the vulnerability.
The result suggests the importance of concealing the defender's recognition and the effectiveness of defensive deception.

\section{Simulation}
\label{sec:sim}

In this section, we confirm the theoretical results through numerical simulation.

\subsection{Fundamental Setup}

\begin{figure}[t]
  \centering
  \includegraphics[width=0.98\linewidth]{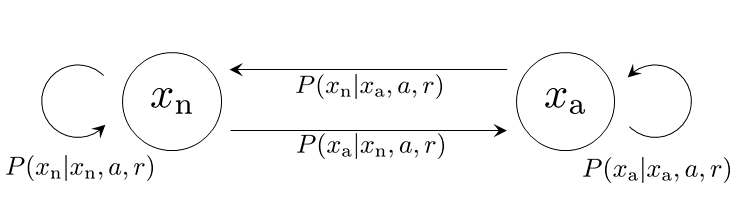}
  \caption{State transition diagram of the numerical example.}
  \label{fig:mot_MDP}
\end{figure}

We assume the state space and the action space to be binary, i.e.,
$\mc{X} = \{x_{\rm n}, x_{\rm a}\}$ and $\mc{A} = \{a_{\rm b},a_{\rm m}\}.$
The states $x_{\rm n}$ and $x_{\rm a}$ represent the normal and abnormal states, respectively,
and $a_{\rm b}$ and $a_{\rm m}$ represent benign and malicious actions, respectively.
The benign and malicious actions correspond to nominal and malicious control signals, respectively.
The reaction set is given by $\mc{R}=\{r_{\rm b},r_{\rm m},r^{\rm b}_{\rm m}\}$.
The state transition diagram is depicted in Fig.~\ref{fig:mot_MDP}.
The initial state is set to $x_{\rm n}$.
The transition probability is given as follows.
Set the transition probability from $x_{\rm n}$ to be given by
\begin{equation}\label{eq:dummy}
 P(x_{\rm a}|x_{\rm n},a,r)=
 \left\{
 \begin{array}{cl}
 0.2 & {\rm if}\ a=a_{\rm b},\\
 0.3 & {\rm if}\ a=a_{\rm m}
 \end{array}
 \right.
\end{equation}
for any $r\in\mc{R}$,
which means that the probability from the normal state to the abnormal state is increased by the malicious action and it is independent of the reaction.
The transition probability from $x_{\rm a}$ to $x_{\rm a}$ is given by Table~\ref{table:prob_xa}.
The probability from the abnormal state to the abnormal state is increased by the malicious action and it is decreased by the reaction $r_{\rm m}$.
The reaction $r^{\rm b}_{\rm m}$ corresponds to bluffing since it induces the same transition probability as $r_{\rm b}$.

\begin{table}[t]
\centering
\caption{Transition Probabilities from the abnormal state to abnormal state.}
\begin{tabular}{c|ccc}
 $P(x_{\rm a}|x_{\rm a},a,r)$ & $r_{\rm b}$ & $r_{\rm m}$ & $r^{\rm b}_{\rm m}$\\ \hline
 $a_{\rm b}$ & 0.5 & 0.3 & 0.5 \\
 $a_{\rm m}$ & 0.6 & 0.4 & 0.6
\end{tabular}
\label{table:prob_xa}
\end{table}

The utilities are given as follows.
The benign sender's utility is
\begin{equation}\label{eq:dummy}
 U^{\rm s}(\tb,x,a,r)=
 \left\{
 \begin{array}{cl}
 1 & {\rm if}\ x=x_n, \\
 0 & {\rm otherwise}
 \end{array}
 \right.
\end{equation}
for any $a\in\mc{A}$ and $r\in\mc{R}$,
which means that the benign sender prefers the nominal state regardless of other variables.
The malicious sender's utility is given by Table~\ref{table:util_mal}.
The benign action $a_{\rm b}$ is a risk-free action, which always induces zero utility, while the malicious action $a_{\rm m}$ is a risky action.
If the reaction is $r_{\rm b},$ the malicious sender obtains positive utility, where the abnormal state $x_{\rm a}$ is more preferred than $x_{\rm n}$.
On the other hand, if the reaction is $r_{\rm m},$ the malicious sender incurs loss.
The receiver's utility is set to be independent on $a\in\mc{A}$ and given by Table~\ref{table:util_rec}, where $a$ is omitted.
The receiver obtains utility only when she takes an appropriate reaction depending on the sender type.
When an appropriate reaction is chosen, the normal state is more preferred than the abnormal state.
Note that $r^{\rm b}_{\rm m}$ induces the same utilities as those with $r_{\rm m}$ but it increases the probability of the abnormal state.
Therefore, there is no motivation to choose $r^{\rm b}_{\rm m}$ when the defender's recognition is known to the attacker.

\begin{table}[t]
\centering
\caption{Malicious Sender's Utility}
\begin{tabular}{c|ccc}
 $U^{\rm s}(\tm,x,a_{\rm b},r)$ & $r_{\rm b}$ & $r_{\rm m}$ & $r^{\rm b}_{\rm m}$\\ \hline
 $x_{\rm n}$ & 0 & 0 & 0 \\
 $x_{\rm a}$ & 0 & 0 & 0
\end{tabular}
\hspace{1mm}
\begin{tabular}{c|ccc}
 $U^{\rm s}(\tm,x,a_{\rm m},r)$ & $r_{\rm b}$ & $r_{\rm m}$ & $r^{\rm b}_{\rm m}$\\ \hline
 $x_{\rm n}$ & 1 & -3 & -3 \\
 $x_{\rm a}$ & 2 & -3 & -3
\end{tabular}
\label{table:util_mal}
\end{table}

\begin{table}[t]
\centering
\caption{Receiver's Utility}
\begin{tabular}{c|ccc}
 $U^{\rm r}(\tb,x,r)$ & $r_{\rm b}$ & $r_{\rm m}$ & $r^{\rm b}_{\rm m}$\\ \hline
 $x_{\rm n}$ & 5 & 0 & 0 \\
 $x_{\rm a}$ & 1 & 0 & 0
\end{tabular}
\hspace{1mm}
\begin{tabular}{c|ccc}
 $U^{\rm r}(\tm,x,r)$ & $r_{\rm b}$ & $r_{\rm m}$ & $r^{\rm b}_{\rm m}$\\ \hline
 $x_{\rm n}$ & 0 & 5 & 5 \\
 $x_{\rm a}$ & 0 & 1 & 1
\end{tabular}
\label{table:util_rec}
\end{table}

Since it is difficult to compute an exact equilibrium for the infinite time horizon problem,
we treat a sequence of equilibria for a finite time horizon problem as a tractable approximation~\cite{Chang2003Two}.
Letting $(\ssen_k,\srec_k,\ldots,\ssen_{k+T-1},\srec_{k+T-1})$ be the resulting equilibrium of the finite time horizon game,
we use $\ssen_k$ and $\srec_k$ as the $k$th strategies as with receding horizon control.
The equilibrium is obtained through brute-force search.
For the game $\mc{G}_2$, the strategies in the simulation are given in a similar manner.
The horizon length is set to $T=2$.
In the numerical examples, the equilibrium is uniquely determined.

\subsection{Simulation: Asymptotic Security}

In the first scenario, we consider the case where the vulnerability is known, and thus this situation corresponds to the game $\mc{G}_1$ in~\eqref{eq:game1}.
The initial belief is given by $\pir_0(\tm)=0.01$, which is known to the sender.
The true sender type is given by $\ts=\tsm$.

\begin{figure}[t]
  \centering
  \includegraphics[width=0.98\linewidth]{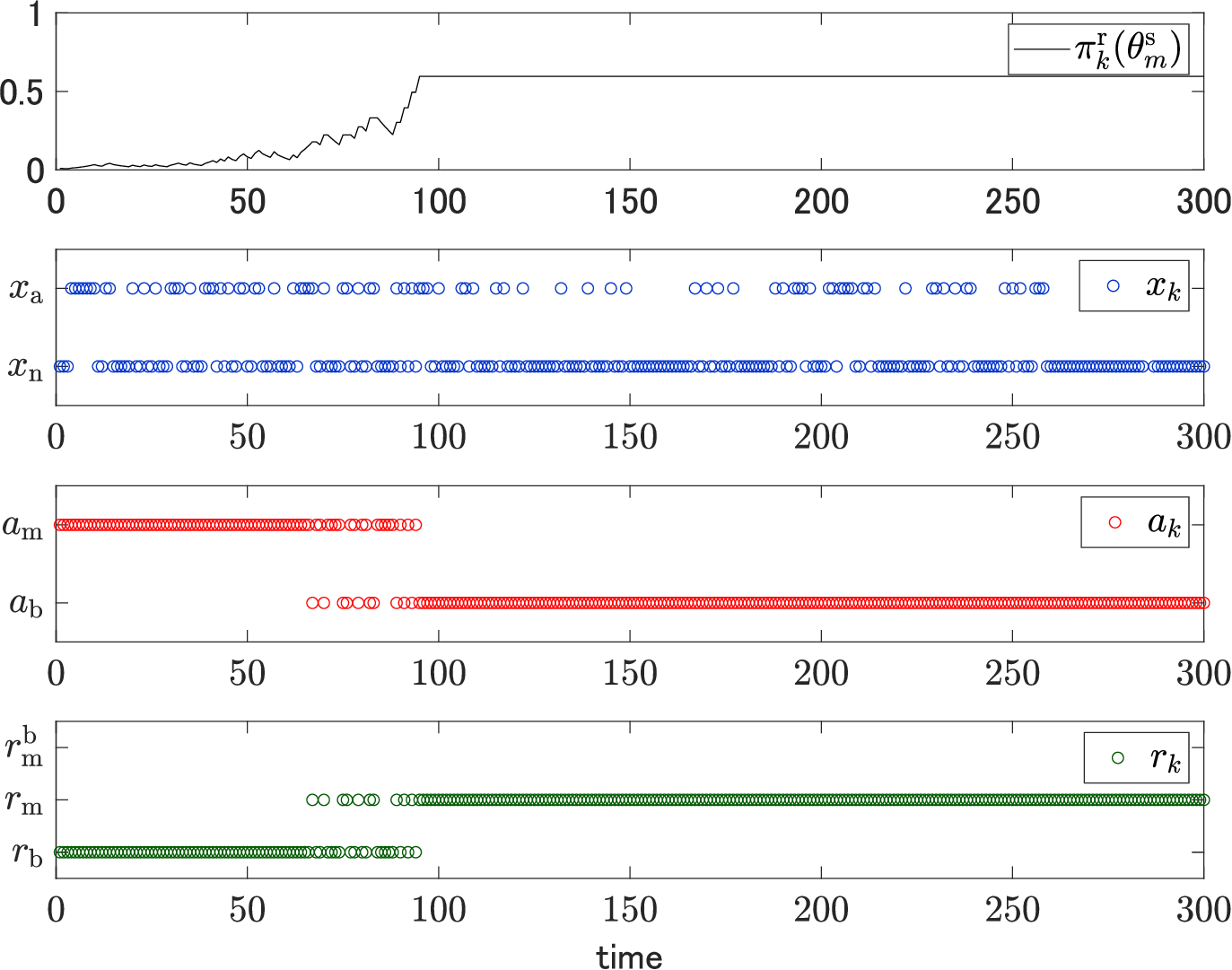}
  \caption{Sample paths of the belief on the malicious sender, state, action, and reactions with $\theta=\tm$.
  The belief converges to a nonzero value over time as claimed by Theorems~\ref{thm:conv} and~\ref{thm:b_zero}.
  The action converges to the benign action as claimed by Theorem~\ref{thm:main_asymp_sec}.
  The results evidence asymptotic security achieved by the Bayesian defense mechanism.
  }
  \label{graph:graph1}
\end{figure}

Under the setting, sample paths of the belief on the malicious sender, the state, the action, and the reaction with $\theta=\tm$ are depicted in Fig.~\ref{graph:graph1}.
The belief converges to a nonzero value over time as claimed by Theorems~\ref{thm:conv} and~\ref{thm:b_zero}.
The action converges to the benign action as claimed by Theorem~\ref{thm:main_asymp_sec}.
The graphs evidence asymptotic security achieved by the Bayesian defense mechanism.
In more detail, it can be observed that the malicious sender takes the malicious action $a_{\rm m}$ while the receiver takes the reaction $r_{\rm b}$ until about the time step $k=50$.
This is because the receiver's belief on the malicious sender is low during the beginning of the game.
On the other hand, between the time steps $k=50$ and $k=100$, $a_{\rm b}$ and $r_{\rm m}$ sporadically appear because the belief is increased.
Finally, after the time step $k=100$, the belief exceeds a threshold, which results in the fixed actions $a=a_{\rm b}$ and $r=r_{\rm m}$.
It is notable that $r^{\rm b}_{\rm m}$ is not chosen at all since there is no reason for it, as explained above.

\subsection{Simulation: Defensive Deception using Bluffing}

\color{black}

In the second scenario, we consider the case where the defender is unaware of the vulnerability and the attacker is unaware of the defender's unawareness.
Then this situation corresponds to the game $\mc{G}_2$ in~\eqref{eq:game2}.
The initial beliefs are given by $\pir_0(\tsm|\tra)=0.3$ and $\pis_0(\tra|\tsm)=0.8.$
The true types are given by $\ts=\tsm$ and $\tr=\tru$.
Note that $\pir_0(\tsm|\tru)=0$ and hence the defender is completely unaware of the attack while the game is proceeding.

We first consider the case where the strategy is \emph{not} passively bluffing.
The same transition probability as that used in the previous simulation, where it depends on the receiver's reaction.
As a result, the state possesses information about the receiver type.

\begin{figure}[t]
  \centering
  \includegraphics[width=0.99\linewidth]{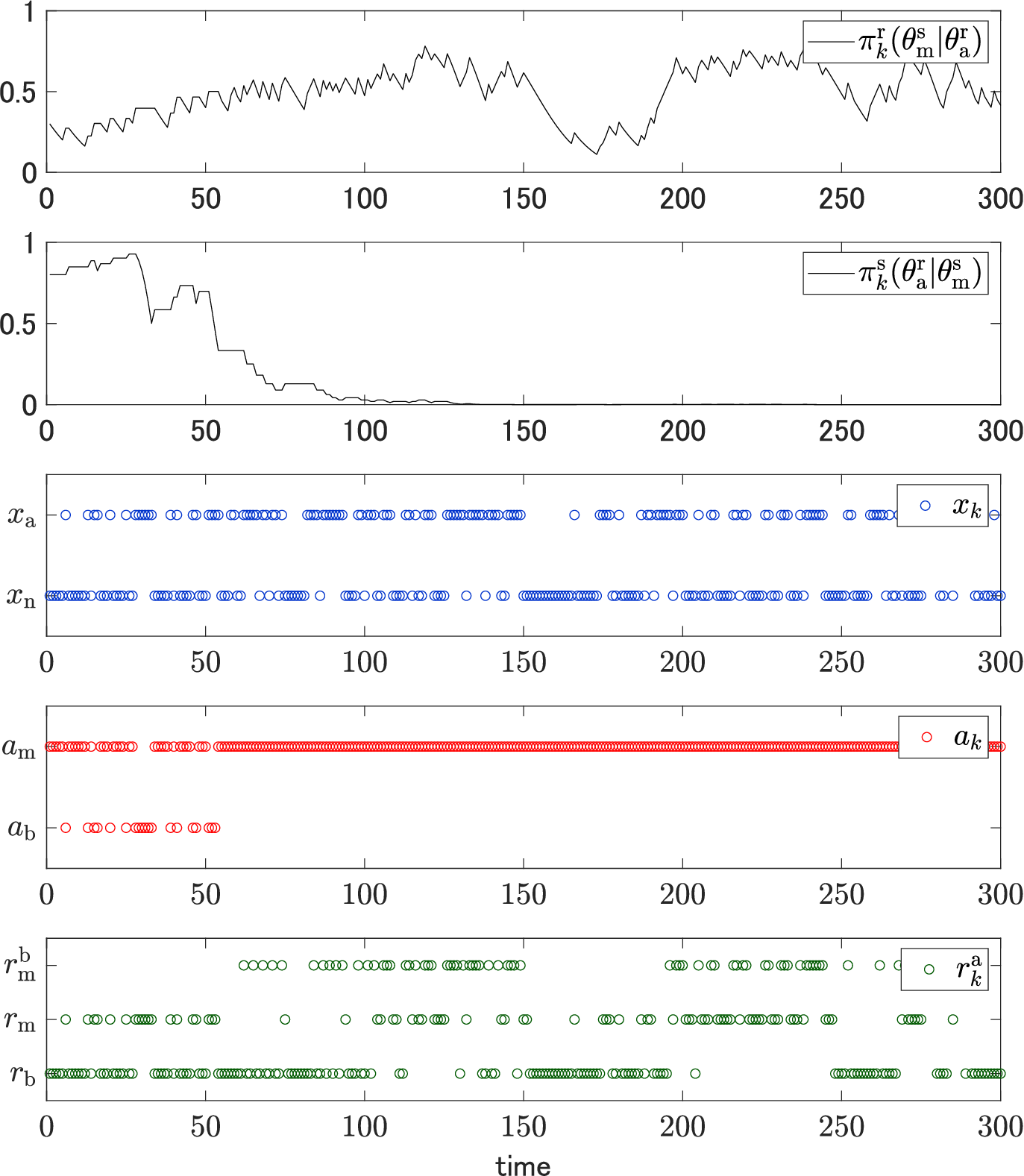}
  \caption{Sample paths of the receiver's belief on the malicious sender when the receiver is aware, the sender's belief on the receiver being aware, state, action,
  and reactions that would be taken when the receiver were aware, where the strategy is \emph{not} passively bluffing.
  The sender's belief converges to zero, i.e., the attacker notices that the defender is unaware of the vulnerability.
  As a result, the malicious action is continuously taken after a sufficiently large number of time steps.
  }
  \label{graph:graph2}
\end{figure}

Fig.~\ref{graph:graph2} depicts sample paths of the receiver's belief on the malicious sender if the receiver were aware of the vulnerability, the sender's belief on the receiver being aware, the actual state, the actual action,
and reactions that would be taken by the receiver being aware.
It can be observed that the sender's belief converges to zero, i.e., the sender notices that the receiver is unaware of the vulnerability.
As a result, malicious actions are constantly taken after a sufficiently large number of time steps.
The result indicates that the defense mechanism fails to defend the system in this case.

We next consider the bluffing case.
As a commitment for passively bluffing strategy, we restrict the reaction set to $\mc{R}=\{r_{\rm b},r^{\rm b}_{\rm m}\}$.
Then the transition probability is independent of the reaction, and hence any strategy becomes passively bluffing.

\begin{figure}[t]
  \centering
  \includegraphics[width=0.98\linewidth]{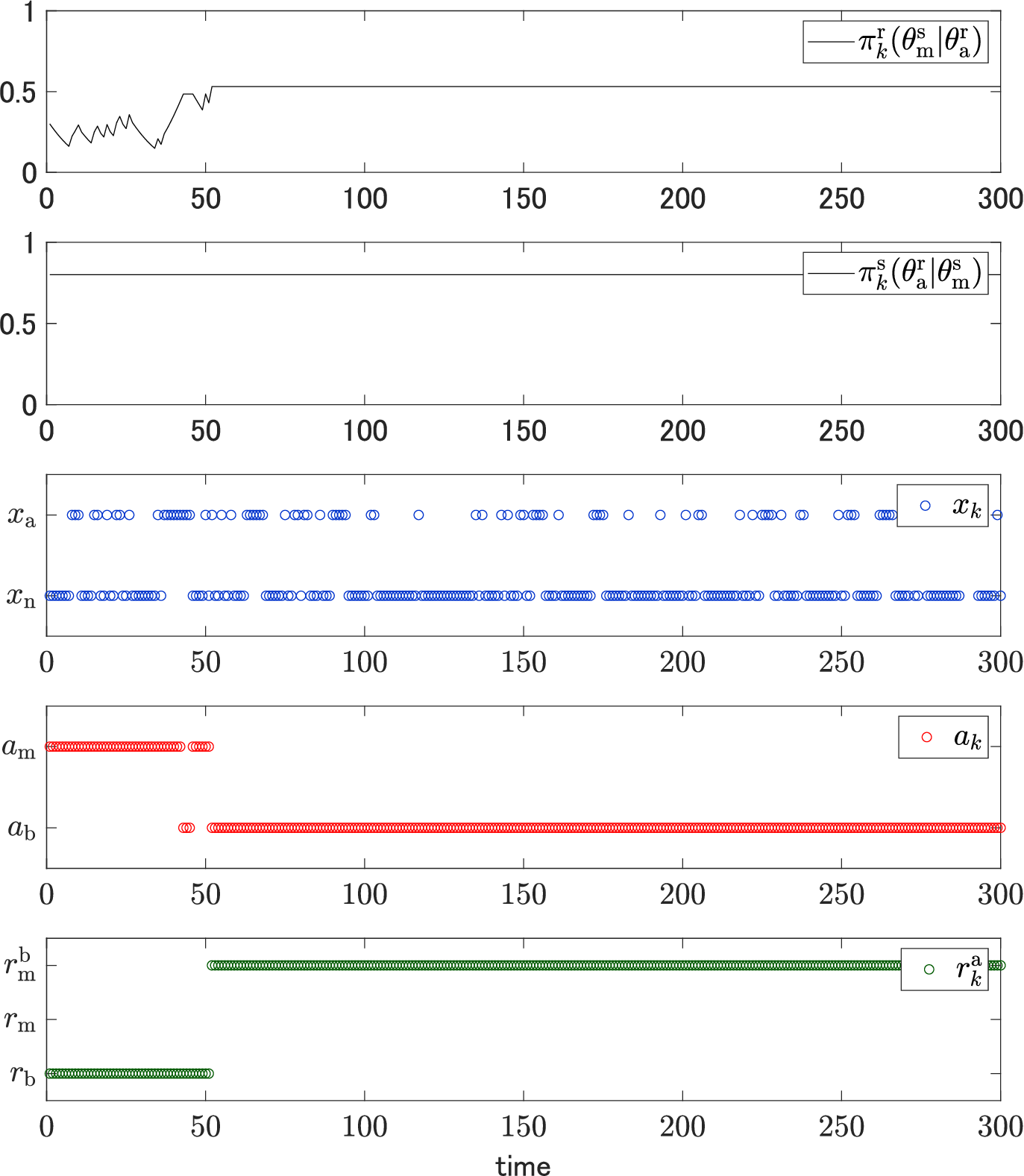}
  \caption{Sample paths of the receiver's belief on the malicious sender when the receiver is aware, the sender's belief on the receiver being aware, state, action,
  and reactions that would be taken when the receiver were aware, where the strategy is passively bluffing.
  The sender's belief is invariant over time because the state does not possess information about the receiver type.
  Thus, the attacker keeps to be cautious about being detected.
  As a result, the benign action is continuously taken after a sufficiently large number of time steps in contrast to Fig.~\ref{graph:graph2}.
  }
  \label{graph:graph3}
\end{figure}

Fig.~\ref{graph:graph3} depicts sample paths of those depicted in Fig.~\ref{graph:graph2} under the bluffing setting.
The sender's belief is invariant over time because the state does not possess information about the receiver type.
Thus, the malicious sender remains cautious about being detected.
As a result, the benign action is continuously taken after a sufficiently large number of time steps in contrast to Fig.~\ref{graph:graph2}.
The result indicates that asymptotic security is achieved by the bluffing even if the defender is unaware of the vulnerability.
The simulation suggests importance of concealing the defender's belief even if it degrades control performance.

\section{Conclusion}
\label{sec:conc}

This study has analyzed defense capability achieved by Bayesian defense mechanisms.
It has been shown that the system to be protected can be guaranteed to be secure by Bayesian defense mechanisms provided that effective countermeasures are implemented.
This fact implies that model knowledge can prevent the defender from being deceived in an asymptotic sense.
As a defensive deception utilizing the derived asymptotic security, bluffing utilizing asymmetric recognition has been considered.
It has also been shown that the attacker possibly stops the execution in the middle of the attack in a rational manner when she strongly believes the defender to be aware of the vulnerability, even if the vulnerability is unnoticed.

Important future work includes an extension to infinite state spaces because the state space in control systems typically is a subset of the Euclidean space.
For this purpose, existing Bayesian consistency analysis for general sample space should be useful~\cite{Walker2004New}.
Moreover, although it is assumed that the state is observable in this framework, a generalization to partially observable setting is a more practical setting.
We expect that the key properties such as Lemma~\ref{lem:submar} still holds if the system requirement, such as Assumption~\ref{assum:input_obs}, can be appropriately modified.
Another direction is to extend the results to non-binary types.
Finally, finding a general condition for detection-averse utilities is an important issue.
For this purpose, the example in Appendix~\ref{app:ex_detection-averse} should be helpful.

\appendices
\section{Example of Detection-averse Utilities}
\label{app:ex_detection-averse}
Consider an MDP with binary spaces $\mc{X}=\{x_{\rm n},x_{\rm a}\}, \mc{A}=\{a_{\rm b},a_{\rm m}\},$ and $\mc{R}=\{r_{\rm b},r_{\rm m}\}$.
The initial state is $x_{\rm n}$ and it goes to $x_{\rm a}$ with probability $p>0$ when $a=a_{\rm m}$ and stays at $x_{\rm m}$ otherwise.
The objective of the receiver is to detect the true state, which is modeled by
\begin{equation}\label{eq:dummy}
 U^{\rm r}(\theta,x,a,r) = \left\{
 \begin{array}{cl}
 1 & {\rm if}\ (\theta,r)=(\tb,r_{\rm b})\ {\rm or}\ (\tm,r_{\rm m}),\\
 0 & {\rm otherwise}.
 \end{array}
 \right.
\end{equation}
The malicious sender's utility is given by
\begin{equation}\label{eq:dummy}
  U^{\rm s}(\tm,x,a,r_{\rm b}) = \left\{
 \begin{array}{cl}
 1 & {\rm if}\ a=a_{\rm m},\\
 0 & {\rm otherwise},
 \end{array}
 \right.
 U^{\rm s}(\tm,x,a,r_{\rm m})=-1,
\end{equation}
which means that the adversary wants to avoid being detected.
The benign sender is assumed to choose $r_{\rm b}$ anytime.
The initial belief satisfies $\pir_0(\tm)<1/2$.
Let $(\ssen,\srec)$ be a strategy such that $\pi^{\tm}_{\infty}=1$ with probability $q>0$.
Take such $x_{0:\infty}$ and then there exists $N\in\Zp$ such that $\pir_k(\tm|x_{0:k})=1$ for any $k>N$ since
\begin{equation}\label{eq:dummy}
 \pir_k(\tm|x_{0:k}) = \left\{
 \begin{array}{ll}
 \pir_0(\tm) & {\rm if}\ x_\tau = x_{\rm n}\ \forall \tau\in\{0,\ldots,k\},\\
 1  & {\rm otherwise}.
 \end{array}
 \right.
\end{equation}
The receiver's best response at $x_{0:k}$ is to take $r_{\rm m}$, which leads to the sender's average utility at $x_{0:k}$ equal $-1$.
Thus, the sender's average utility at the initial state is $-q<0$.
However, if the sender takes the strategy such that $a_{\rm k}=a_{\rm b}$ for any $k\in\Zp$, then the sender's average utility at the initial state is $0$.
Thus, $\ssen$ is not a best response to $\srec$, which means that the utilities are detection-averse.

\section{Proofs of Propositions}
In the proofs, we omit the symbol $s$ in the notation for simplicity when no confusion arises.

\begin{proof}
{\it Proof of Lemma~\ref{lem:submar}:}
It is clear that $\pi^\theta_k$ is adapted to the filtration $\sigma(X_{0:k})$.
It is also clear that $\pi^\theta_k$ is integrable with respect to $\Prob_{\theta}$ since it is bounded.
Thus, it suffices to show
\begin{equation}\label{eq:dummy}
 \Exp\left[\pi^\theta_{k+1}|\sigma(X_{0:k})\right]\geq \pi^\theta_k\quad \Prob_{\theta}{\rm -a.s.}
\end{equation}
for the claim.
For a fixed outcome $\omega\in\Omega$ with which $X_{0:k}(\omega)=x_{0:k}$, the inequality is equivalent to
\begin{equation}\label{eq:submar}
 \sum_{x_{k+1}\in\mc{X}} p_\theta(x_{k+1}|x_{0:k}) \pi_{k+1}(\theta|x_{0:k+1})\geq \pir_k(\theta|x_{0:k}).
\end{equation}
Thus it suffices to show~\eqref{eq:submar} for any $k\in\NN$ and $x_{0:k}\in\mc{X}^{k+1}$.

First, we reduce the index of the summation in~\eqref{eq:submar}.
When $\pir_k(\theta|x_{0:k})=0$, the inequality~\eqref{eq:submar} always holds.
Thus we assume $\pir_k(\theta|x_{0:k})>0$ in the following.
Define
\begin{equation}\label{eq:dummy}
 \mc{X}^0_k:=\left\{x_{k+1}\in\mc{X}: \sum_{\phi\in\iTheta}p_\phi(x_{k+1}|x_{0:k})\pir_k(\phi|x_{0:k})=0\right\}.
\end{equation}
Because $\pir_k(\phi|x_{0:k})$ is positive for any $\phi\in\iTheta$, if $x_{k+1}$ belongs to $\mc{X}^0_k$ then $p_{\theta}(x_{k+1}|x_{0:k})=0$.
Hence~\eqref{eq:submar} is equivalent to
\begin{equation}\label{eq:submar2}
 \sum_{x_{k+1}\in\mc{X}^+_k} p_{\theta}(x_{k+1}|x_{0:k}) \pir_{k+1}(\theta|x_{0:k+1}) \geq \pir_k(\theta|x_{0:k})
\end{equation}
where $\mc{X}^+_k:=\mc{X}\setminus\mc{X}^0_k$.

To simplify notation, we define
\begin{equation}\label{eq:notation}
 \pi(\theta):=\pir_k(\theta|x_{0:k}),\ p_\phi(x):=p_{\phi}(x|x_{0:k}),\ \mc{X}^+:=\mc{X}^+_k
\end{equation}
for fixed $k$ and $x_{0:k}$.
Then the inequality~\eqref{eq:submar2} is equivalent to
\begin{equation}\label{eq:dummy}
 \sum_{x\in\mc{X}^+} p_{\theta}(x)
 \dfrac{p_\theta(x) \pi(\theta) }
 {\sum_{\phi\in\iTheta} p_\phi(x) \pi(\phi) }
 \geq \pi(\theta).
\end{equation}
Because $\pi(\theta)>0$, this inequality is equivalent to
\begin{equation}\label{eq:submar3}
 \underbrace{
 \sum_{x\in\mc{X}^+}
 p_{\theta}(x)
 \dfrac{ p_{\theta}(x)}
 {\sum_{\phi\in\iTheta} p_\phi(x) \pi(\phi) }
 }_{=:G(\theta)}
 \geq 1.
\end{equation}
By rewriting $G(\theta)$, we have
\begin{equation}\label{eq:dummy}
 \begin{array}{cl}
 G(\theta) \hs = \displaystyle{
 \sum_{x\in\mc{X}^+}}
 p_\theta(x)
 \dfrac{1}
 {
 \sum_{\phi\in\iTheta} \frac{p_\phi(x)}{p_\theta(x)}\pi(\phi)
 }.
 \end{array}
\end{equation}
By applying Jensen's inequality in~\eqref{eq:Jensen} with the functions $p(x):=p_\theta(x),a(x):=\sum_{\phi\in\iTheta} \frac{p_\phi(x)}{p_\theta(x)}\pi(\phi),$ and $\varphi(\xi):=1/\xi$, we have
\begin{equation}\label{eq:dummy}
\begin{array}{cl}
 G(\theta) \hs \geq 
 \varphi\left(\sum_{x\in\mc{X}^+} p_\theta(x) \sum_{\phi\in\iTheta} \frac{p_\phi(x)}{p_\theta(x)} \pi(\phi) \right)\\
 \hs = \varphi\left(\sum_{x\in\mc{X}^+} \sum_{\phi\in\iTheta} p_\phi(x) \pi(\phi) \right)\\
 \hs = \varphi\left(\sum_{\phi\in\iTheta} \left(\sum_{x\in\mc{X}^+} p_\phi(x)\right)\pi(\phi)  \right)\\
 \hs = \varphi\left(\sum_{\phi\in\iTheta} \pi(\phi) \right)\\
 \hs = \varphi(1)\\
 \hs = 1,
\end{array}
\end{equation}
which leads to the claim.
\end{proof}

\begin{proof}
{\it Proof of Theorem~\ref{thm:conv}:}
Because the belief is uniformly bounded over time, we have $\sup_{k\in\NN} \Exp_\theta\left[\pi^\theta_k\right]<\infty.$
From Lemma~\ref{lem:submar} and Doob's convergence theorem~\cite[Theorem~4.4.1]{Cinlar2011Probability}, the claim holds.
\end{proof}

\begin{proof}
{\it Proof of Lemma~\ref{lem:log_submar}:}
Since $\pi^\theta_0>0$, we have $\pi^\theta_k>0$ $\Prob_\theta$-almost surely for any $k\in\NN$.
Thus $\log(\pi^\theta_k)$ is well-defined.
We first show that $\log(\pi^\theta_k)$ is a submartingale with respect to the probability measure $\Prob_\theta$ and the filtration $\sigma(X_{0:k})$.
It is clear that $\log(\pi^\theta_k)$ is adapted to the filtration $\sigma(X_{0:k})$.
Because the number of elements in the support of $\log(\pi^\theta_k)$ is finite, $\log(\pi^\theta_k)$ is integrable for any $k\in\NN$.
Thus it suffices to show that
\begin{equation}\label{eq:dummy}
 \Exp_\theta\left[\log(\pi^\theta_{k+1})|\sigma(X_{0:k})\right]\geq \log(\pi^\theta_k)\quad \Prob_\theta{\rm -a.s.}
\end{equation}
As in the proof of Lemma~\ref{lem:submar}, this inequality is equivalent to
\begin{equation}\label{eq:dummy}
 \sum_{x_{k+1}\in\mc{X}^+_k} p_{\theta}(x_{k+1}|x_{0:k}) \log(\pir_{k+1}(\theta|x_{0:k+1})) \geq \log(\pir_k(\theta|x_{0:k}))
\end{equation}
for any $x_{0:k}\in\mc{X}^{k+1}$.
With the notation~\eqref{eq:notation}, this is equivalent to
\begin{equation}\label{eq:Mar_in_log}
 \sum_{x\in\mc{X}^+} p_{\theta}(x)
 \log\left(\dfrac{p_\theta(x) \pi(\theta) }
 {\sum_{\phi\in\iTheta} p_\phi(x) \pi(\phi)} \right)
 \geq \log(\pi(\theta)).
\end{equation}
Because the left-hand side can be rewritten by
\begin{equation}\label{eq:dummy}
 \begin{array}{cl}
  & \displaystyle{\sum_{x\in\mc{X}^+} p_{\theta}(x)}
 \log\left(\dfrac{p_\theta(x) \pi(\theta) }
 {\sum_{\phi\in\iTheta} p_\phi(x) \pi(\phi)} \right)\\
 = & \displaystyle{\sum_{x\in\mc{X}^+} p_{\theta}(x)}
 \left\{
 \log\left(\dfrac{p_\theta(x) }
 {\sum_{\phi\in\iTheta} p_\phi(x) \pi(\phi)} \right)
 + \log(\pi(\theta)) \right\}\\
 = & \displaystyle{\sum_{x\in\mc{X}^+} p_{\theta}(x)}
 \log\left(\dfrac{p_\theta(x) }
 {\sum_{\phi\in\iTheta} p_\phi(x) \pi(\phi)} \right)
 + \log(\pi(\theta)),
 \end{array}
\end{equation}
the inequality~\eqref{eq:Mar_in_log} is equivalent to
\begin{equation}\label{eq:dummy}
 \sum_{x\in\mc{X}^+} p_{\theta}(x)
 \log\left(\dfrac{p_\theta(x) }
 {\sum_{\phi\in\iTheta} p_\phi(x) \pi(\phi)} \right)
 \geq 0,
\end{equation}
which is also equivalent to
\begin{equation}\label{eq:dummy}
 \underbrace{
 \sum_{x\in\mc{X}^+} p_{\theta}(x)
 \log\left(\dfrac{
 \sum_{\phi\in\iTheta} p_\phi(x) \pi(\phi)}{p_\theta(x)}
 \right)
 }_{H(\theta)}
 \leq 0.
\end{equation}
By applying Jensen's inequality for a concave function with
$p(x):=p_\theta(x),a(x):=\frac{\sum_{\phi\in\iTheta}p_\phi(x)\pi(\phi)}{p_\theta(x)},$ and $\varphi(\xi):=\log(\xi),$ we have
\begin{equation}\label{eq:dummy}
\begin{array}{cl}
 H(\theta) \hs \leq \log\left( \sum_{x\in\mc{X}^+} p_\theta(x)
 \frac{\sum_{\phi\in\iTheta}p_\phi(x)\pi(\phi)}{p_\theta(x)}
 \right)\\
 \hs = \log\left( \sum_{x\in\mc{X}^+} \sum_{\phi\in\iTheta} p_\phi(x)\pi(\phi)\right)\\
 \hs = \log(1)\\
 \hs = 0,
\end{array}
\end{equation}
which implies that $\log(\pi^\theta_k)$ is a submartingale.

From Doob's convergence theorem~\cite[Theorem~4.4.1]{Cinlar2011Probability}, it suffices to show that the expectation of the nonnegative part of $\log(\pi^\theta_k)$ is uniformly bounded.
Because $\pi^\theta_k\in(0,1]$, $\log(\pi^\theta_k)$ is nonpositive for any $k\in\NN$, and hence the uniform boundedness holds.
\end{proof}

\begin{proof}
{\it Proof of Theorem~\ref{thm:b_zero}:}
We prove the claim by contradiction.
Define $E$ as the inverse image of $\{0\}$ for $\pi^\theta_{\infty}$.
Assume that $\Prob_\theta(E)>0$.
For any $\omega\in E,$ $\pi^\theta_k(\omega)\to 0$ as $k\to \infty$.
Hence, from the continuity of logarithm functions, it turns out that $\log(\pi^\theta_k(\omega))\to -\infty$ as $k\to \infty$.
This means that $\log(\pi^\theta_k(\omega))$ diverges for $\omega\in E$.
However, Lemma~\ref{lem:log_submar} states that $\log(\pi^\theta_k)$ converges $\Prob_\theta$-almost surely.
This is a contradiction.
\end{proof}

\begin{proof}
{\it Proof of Lemma~\ref{lem:pp}:}
We first show that the coefficient of Bayes' rule converges to one, i.e.,
\begin{equation}\label{eq:f_conv}
 \lim_{k\to\infty}f_k(\tm,X_{0:k})=1\quad \Prob_{\tm}{\rm -a.s.}
\end{equation}
where $f_{k+1}(\theta,x_{0:k}):=\pi^{\theta}_{k+1}/\pi^{\theta}_k$.
Because $\pi^{\tm}_\infty$ is nonzero $\Prob_{\tm}$-almost surely from Theorem~\ref{thm:b_zero}, we have
\begin{equation}\label{eq:dummy}
 \begin{array}{cl}
  \displaystyle{ \lim_{k\to\infty} f_{k+1}(\tm,X_{0:k})} \hs \displaystyle{ =\lim_{k\to\infty} \left(\pi^{\tm}_{k+1}
  /\pi^{\tm}_k\right)}\\
 \hs  = \pi^{\tm}_\infty/\pi^{\tm}_\infty\\
 \hs = 1
 \end{array}
\end{equation}
$\Prob_{\tm}$-almost surely.
Thus~\eqref{eq:f_conv} holds.

It is observed that $f_k$ can be calculated by
\begin{equation}\label{eq:dummy}
 f^s_{k}(\theta,x_{0:k}) = \dfrac{p^s_{\theta} (x_{k}|x_{0:k-1})}{\sum_{\phi\in\iTheta}p^s_\phi (x_{k}|x_{0:k-1}) \pir_{k-1}(\phi|x_{0:k-1})}
\end{equation}
from~\eqref{eq:Bayes}.
Define
\begin{equation}\label{eq:dummy}
\begin{array}{l}
 f_{{\rm N},k+1}(\omega):= p_{\tm}(X_{k+1}(\omega)|X_{0:k}(\omega)),\\
 f_{{\rm D},k+1}(\omega):= \sum_{\phi\in\iTheta}p_{\phi}(X_{k+1}(\omega)|X_{0:k}(\omega))\pi^{\phi}_k(\omega),
\end{array}
\end{equation}
which denote the numerator and the denominator of the coefficient $f_{k+1}(\tm,X_{0:k+1}(\omega))$, respectively.
Since $\pi^{\tm}_0>0$, we have $0<f_{{\rm D},k+1}\leq 1$ for any $\omega\in\Omega,k\in\Zp$.
Thus
\begin{equation}\label{eq:squeeze_ineq}
 0 \leq f_{{\rm D},k+1}|f_{k+1}-1|\leq |f_{k+1}-1|\quad \Prob_{\tm}{\rm -a.s.}
\end{equation}
for any $k\in\Zp$.
Because
$\lim_{k\to\infty}|f_{k+1}-1|= 0\quad \Prob_{\tm}{\rm -a.s.}$
from~\eqref{eq:f_conv}, the squeeze theorem for~\eqref{eq:squeeze_ineq} yields
\begin{equation}\label{eq:ffz}
 \lim_{k\to\infty}f_{{\rm D},k+1}|f^{\tm,s}_{k+1}-1|= 0\quad \Prob_{\tm}{\rm -a.s.}
\end{equation}
Now we have
\begin{equation}\label{eq:dummy}
 \begin{array}{l}
  f_{{\rm D},k+1}|f^{\tm,s}_{k+1}-1|\\
 =  |f_{{\rm N},k+1} - f_{{\rm D},k+1}|\\
 =  |
 p_{\tm}(X_{k+1}|X_{0:k}) - \sum_{\phi\in\iTheta} p_\theta(X_{k+1}|X_{0:k})\pi^\theta_k
 |\\
 = |p_{\tm}(X_{k+1}|X_{0:k})(1-\pi^{\tm}_k) - p_{\tb}(X_{k+1}|X_{0:k})\pi^{\tb}_k|\\
 = |p_{\tm}(X_{k+1}|X_{0:k}) - p_{\tb}(X_{k+1}|X_{0:k})|(1-\pi^{\tm}_k).
 \end{array}
\end{equation}
Since $s$ is a PBE with detection-averse utilities,
$\lim_{k\to\infty}(1-\pi^{\tm}_k)\neq 0\quad \Prob_{\tm}{\rm -a.s.}$
Therefore,~\eqref{eq:ffz} leads to the claim.
\end{proof}

\begin{proof}
{\it Proof of Theorem~\ref{thm:main_asymp_sec}:}
From the definition of the conditional probability mass function, we have
\begin{equation}\label{eq:dummy}
 p_{\theta}(X_{k+1}|X_{0:k}) = P(X_{k+1}|X_k,\ssen_k(\theta,X_{0:k}),\srec_k(X_{0:k})).
\end{equation}
Thus the claim of Lemma~\ref{lem:pp} can be rewritten by
\begin{equation}\label{eq:P_limit}
\begin{array}{l}
 | P(X_{k+1}|X_k,A^{\tm}_k,\srec_k(X_{0:k}))  - P(X_{k+1}|X_k,A^{\tb}_k,\srec_k(X_{0:k})) |\\
  \to 0
\end{array}
\end{equation}
$\Prob_{\tm}$-almost surely as $k\to\infty$.
From finiteness of the MDP, the condition~\eqref{eq:P_limit} is equivalent to
\begin{equation}\label{eq:Ek_ioz}
 \Prob_{\tm}(\{E_k\ \io\})=0
\end{equation}
where
\begin{equation}\label{eq:dummy}
 E_k:=  \left\{ P(X_{k+1}|X_k,A^{\tm}_k,R_k) \neq P(X_{k+1}|X_k,A^{\tb}_k,R_k)\right\}.
\end{equation}
By applying the generalized Borel-Cantelli's second lemma in~\eqref{eq:Borel-Cantelli} with $\mc{F}_k:=\sigma(X_{0:k}),$ we have
\begin{equation}\label{eq:PEz}
 \Prob_{\tm}(E)=0
\end{equation}
where
\begin{equation}\label{eq:dummy}
\textstyle{
 E:= \left\{ \omega\in\Omega:
 \sum_{k=0}^\infty \Prob_{\tm} \left( E_k|\sigma(X_{0:k})\right) (\omega)=\infty
 \right\}.
}
\end{equation}

We derive a simpler description of the event $E$.
For any $\omega\in \Omega$, the set of nonnegative integers $\Zp$ can be divided into two disjoint subsets $\hat{\mathbb{Z}}_+(\omega)$ and $\Zp \setminus \hat{\mathbb{Z}}_+(\omega)$ such that
\begin{equation}\label{eq:dummy}
 \left\{
 \begin{array}{l}
 A^{\tm}_{k}(\omega) \neq A^{\tb}_{k}(\omega),\quad \forall k\in \hat{\mathbb{Z}}_+(\omega), \\
 A^{\tm}_{k}(\omega) = A^{\tb}_{k}(\omega),\quad \forall k\in \Zp \setminus \hat{\mathbb{Z}}_+(\omega).
 \end{array}
 \right.
\end{equation}
For a fixed $\omega\in\Omega$, $\Prob_{\tm}(E_k|\sigma(X_{0:k}))(\omega)=0$ for $k\in \Zp \setminus \hat{\mathbb{Z}}_+(\omega)$.
Thus we have
\begin{equation}\label{eq:dummy}
 \sum_{k=0}^\infty \Prob_{\tm} \left( E_k|\sigma(X_{0:k})\right) (\omega) = \sum_{k\in \hat{\mathbb{Z}}_+(\omega)} \Prob_{\tm} \left( E_k|\sigma(X_{0:k})\right) (\omega).
\end{equation}
Moreover, for any $k\in\Zp$ and $\omega\in\Omega$, we have
\begin{equation}\label{eq:dummy}
 \Prob_{\tm} \left( E_k|\sigma(X_{0:k})\right) (\omega) = \sum_{x_{k+1}\in\mc{X}_{k+1}(\omega)} P(x_{k+1}|x_k,a^{\tm}_k,r_k)
\end{equation}
where
\begin{equation}\label{eq:dummy}
 \mc{X}_{k+1}(\omega):=\{x\in\mc{X}: P(x|x_k,a^{\tm}_k,r_k) \neq  P(x|x_k,a^{\tb}_k,r_k) \}
\end{equation}
with $x_{0:k}:=X_{0:k}(\omega),a^{\theta}_k:=A^{\theta}_k(\omega),$ and $r_k:=R_k(\omega)$.
Thus,
the condition in the definition of $E$ can be rewritten by
\begin{equation}\label{eq:E_con}
 \displaystyle{
 \sum_{k\in\hat{\mathbb{Z}}_+(\omega)} \sum_{x_{k+1}\in\mc{X}_{k+1}(\omega)} P(x_{k+1}|x_k,a^{\tm}_k,r_k)=\infty.
 } 
\end{equation}

Now we define
$F:= \{ A^{\tm}_k \neq A^{\tb}_k\ \io \}.$
We show the claim
by contradiction.
Assume $\Prob_{\tm}(F)>0$.
Then $\Prob_{\tm}(E|F)$ is well-defined.
From~\eqref{eq:PEz}, we have $\Prob_{\tm}(E\cap F)=\Prob_{\tm}(E|F)\Prob_{\tm}(F)=0.$
Because $\Prob_{\tm}(F)$ is assumed to be nonzero, this equation implies
$\Prob_{\tm}(E|F)=0.$
We now calculate $\Prob_{\tm}(E|F)$ from its definition.
Let
\begin{equation}\label{eq:dummy}
 \gamma(\omega):= \inf_{k\in\hat{Z}_+(\omega)} \sum_{x_{k+1}\in\mc{X}_{k+1}(\omega)} P(x_{k+1}|x_k,a^{\tm}_k,r_k).
\end{equation}
For any $\omega\in \Omega$, the set $\mc{X}_{k+1}(\omega)$ is nonempty for $k\in\hat{\mathbb{Z}}_+(\omega)$ from Assumption~\ref{assum:input_obs}.
This fact and the finiteness of the MDP lead to that $\gamma(\omega) >0.$
The infimum leads to the inequality
\begin{equation}\label{eq:Pgam}
 \sum_{k\in\hat{\mathbb{Z}}_+(\omega)} \sum_{x_{k+1}\in\mc{X}_{k+1}(\omega)} P(x_{k+1}|x_k,a^{\tm}_k,r_k) \geq \left|\hat{\mathbb{Z}}_+(\omega)\right| \gamma(\omega).
\end{equation}
If $\omega\in F$, then $\hat{\mathbb{Z}}_+(\omega)$ has infinite elements and hence $|\hat{\mathbb{Z}}_+(\omega)|\gamma(\omega)=\infty$.
Thus, for any $\omega\in F$, from the inequality~\eqref{eq:Pgam}, the condition~\eqref{eq:E_con} holds.
Therefore, $\Prob_{\tm}(E|F)=1$, which is a contradiction.
Hence $\Prob_{\tm}(F)=0$ holds.
\end{proof}

\begin{proof}
{\it Proof of Lemma~\ref{lem:G1G2_eq}:}
The former claim is obvious since the probability measures are independent of the strategies with $\tsb$ and $\tru$.
Assume $\hat{s}^{\rm s}_{2,k:\infty} \in \BRs_k(\hat{s}^{\rm r}_{2,k:\infty}|\tsm,x_{0:k})$ for any $k\in\Zp$ and $x_{0:k}\in\mc{X}^{k+1}$.
Because $\hpis_k(\tra|\tsm)=1$ and $\hpis_k(\tru|\tsm)=0$ for $k\in\Zp$, the malicious sender's expected average utility in $\hat{\mc{G}}_2$ is given by
\begin{equation}\label{eq:dummy}
 \begin{array}{l}
\bar{U}^{\rm s}_{k,T}(\hat{s}_{2,{k:k+T}}|\tsm,x_{0:k}):= \dfrac{1}{T+1}\\
\displaystyle{\small{
  \times\Exp^s\left[\left. \sum_{\tau=k}^{k+T} U^{\rm s}(\tsm,X_k,\hat{s}^{\rm s}_{2,\tau}(\tsm,X_{0:\tau}),\hat{s}^{\rm r}_{2,\tau}(\tra,X_{0:\tau})) \right| x_{0:k}\right].
  }
}
\end{array}
\end{equation}
which is the malicious sender's utility in $\mc{G}_1$.
Thus, $\BRs_k(\hat{s}^{\rm r}_{2,k:\infty}|\tsm,x_{0:k})$ in $\hat{\mc{G}}_2$ is equal to $\BRs_k(s^{\rm r}_{1,k:\infty}|\tm,x_{0:k})$ in $\mc{G}_1$.
Hence, $\ssen_{1,k:\infty}\in \BRs_k(\srec_{1,k:\infty}|\tm,x_{0:k}).$
\end{proof}

\begin{proof}
{\it Proof of Lemma~\ref{lem:relax_asymp_benign}:}
If $s$ is a passively bluffing strategy, then the distribution of $H^{\rm s}_k$ is independent of the receiver type.
Thus the distribution of $\delta\left(A^{\tm}_k,A^{\tb}_k\right)$ is also independent of the receiver type.
Hence, if~\eqref{eq:asymp_benign_tra} holds, the same condition holds for $\tru$ as well.
\end{proof}

\begin{proof}
{\it Proof of Lemma~\ref{lem:rephrase}:}
Take $s=(\ssen,\srec)\in\mc{S}_{\rm nab}^{\ast}$.
Consider $\mc{G}_1$ corresponding to $\hat{\mc{G}}_2$.
Let $s_1=(\ssen_1,\srec_1)$ be a strategy profile in $\mc{G}_1$ given by~\eqref{eq:str_restriction}.
From Lemma~\ref{lem:G1G2_eq}, we have
\begin{equation}\label{eq:dummy}
 \Prob^{s_1}_{\tsm} \left( \delta( A^{\tm}_k, A^{\tb}_k )=0\right) = \Prob^{\hat{s}_2}_{\tsm,\tra} \left( \delta( A^{\tm}_k, A^{\tb}_k )=0\right).
\end{equation}
From the contraposition of Lemma~\ref{lem:relax_asymp_benign}, this equation implies that $s_1$ is not asymptotically benign in the sense of the game $\mc{G}_1$ since $s\in\mc{S}_{\rm nab}$.
Thus, $s_1$ is not a PBE of $\mc{G}_1$ from Theorem~\ref{thm:main_asymp_sec}.
This means that $\ssen_1$ contains a strategy that is not a best response.
Because $s\in\mc{S}^{\ast}$, this means $\ssen_{1,k:\infty}\not \in \BRs_k(\srec_{1,k:\infty}|\tm,x_{0:k})$ for some $k$.
From the contraposition of Lemma~\ref{lem:G1G2_eq}, $\ssen_{k:\infty}\not\in \BRs_k(\srec_{k:\infty}|\tsm,x_{0:k})$, which is equivalent to~\eqref{eq:non_PBE}.
\end{proof}

\begin{proof}
{\it Proof of Theorem~\ref{thm:bluffing}:}
We prove the existence of $\pis_0(\tra|\tsm)<1$ such that the contraposition of the condition holds, i.e., if $s$ is not asymptotically benign then $s$ is not a PBE.
Let $s\in\mc{S}_{\rm nab}$.
If $s\not\in\mc{S}^{\ast}$, $s$ is not a PBE.
Thus we suppose $s\in\mc{S}_{\rm nab}^{\ast}$.
It suffices to show that there exists $\pis_0(\tra|\tsm)<1$ such that
\begin{equation}\label{eq:supsD}
 \inf_{s\in \mc{S}_{\rm nab}^{\ast}} D(s,g(s))>0
\end{equation}
where
$D(s,\tssen) := \bar{U}^{\rm s}((\tssen,\srec),\tsm,\pis)-\bar{U}^{\rm s}(s,\tsm,\pis)$
and $g$ is given in~\eqref{eq:func_g}.

From~\eqref{eq:bUs_weighted_sum}, we have
$D(s,\tssen) = \sum_{\tr\in\iThetar}\pis_0(\tr|\tsm).$
From the definition of $\gamma$ in~\eqref{eq:inf_sup}, we have
\begin{equation}\label{eq:D_ineq}
 \begin{array}{cl}
   D(s,g(s)) \hs \geq D_{\tru}(s,g(s)) \pis_0(\tru|\tsm) + \gamma\pis_0(\tra|\tsm)\\
 \hs = \gamma + (D_{\tru}(s,g(s))-\gamma)\pis_0(\tru|\tsm)
 \end{array}
\end{equation}
Consider the case where $D_{\tru}(s,g(s))-\gamma\geq 0$ for any $s\in \mc{S}_{\rm nab}^{\ast}$.
Then~\eqref{eq:D_ineq} implies that $D(s,g(s))\geq \gamma$  for any $\pis_0(\tra|\tsm)<1$.
From Assumption~\ref{assum:uniformly_bounded}, this inequality leads to
$\inf_{s\in \mc{S}_{\rm nab}^{\ast}} D(s,g(s)) \geq \gamma > 0,$
which implies~\eqref{eq:supsD}.

Next, consider the case where $D_{\tru}(s,g(s))-\gamma<0$ for some $s\in \mc{S}_{\rm nab}^{\ast}$.
By taking an initial belief $\pis_0(\tru|\tsm)>0$ such that
\begin{equation}\label{eq:pi_cond}
 \pis_0(\tru|\tsm) < \inf_{s \in \mc{T}} \dfrac{-\gamma}{D_{\tru}(s,g(s))-\gamma}
\end{equation}
where
$\mc{T}:= \{ s\in \mc{S}_{\rm nab}^{\ast}: D_{\tru}(s,g(s))-\gamma<0 \},$
we have~\eqref{eq:supsD} from~\eqref{eq:D_ineq}.
From the definition of $D_{\tru}$ in~\eqref{eq:D_tr}, $D_{\tru}(s,g(s))$ is bounded in $\mc{T}$ because $\bUs_{\tr}$ is bounded for any strategy.
Thus we have
\begin{equation}\label{eq:dummy}
 \inf_{s \in \mc{T}} \dfrac{-\gamma}{D_{\tru}(s,g(s))-\gamma}>0.
\end{equation}
Thus a nonzero initial belief that satisfies~\eqref{eq:pi_cond} exists.

\end{proof}

\bibliography{sshrrefs}
\bibliographystyle{IEEEtran}

\begin{IEEEbiography}[{\includegraphics[width=1in,height=1.25in,clip,keepaspectratio]{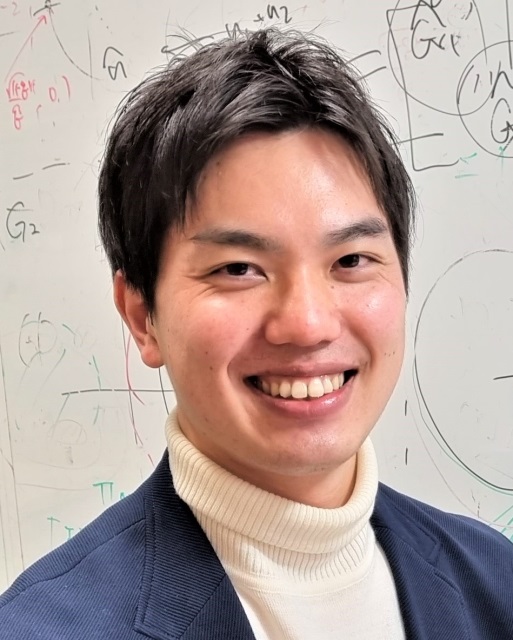}}]{Hampei Sasahara}(M'15)
received the Ph.D. degree in engineering from Tokyo Institute of Technology in 2019.
He is currently an Assistant Professor with Tokyo Institute of Technology, Tokyo, Japan.
From 2019 to 2021, he was a Postdoctoral Scholar with KTH Royal Institute of Technology, Stockholm, Sweden.
His main interests include secure control system design and control of large-scale systems.
\end{IEEEbiography}

\begin{IEEEbiography}[{\includegraphics[width=1in,height=1.25in,clip,keepaspectratio]{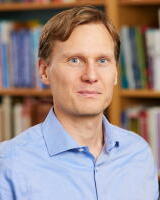}}]{Henrik Sandberg}(F'23)
received the M.Sc. degree in engineering physics and the Ph.D. degree in automatic control from Lund University, Lund, Sweden, in 1999 and 2004, respectively.
He is a Professor with the Division of Decision and Control Systems, KTH Royal Institute of Technology, Stockholm, Sweden.
From 2005 to 2007, he was a Postdoctoral Scholar with the California Institute of Technology, Pasadena, CA, USA.
In 2013, he was a Visiting Scholar with the Laboratory for Information and Decision Systems, Massachusetts Institute of Technology, Cambridge, MA, USA.
He has also held visiting appointments with the Australian National University, Canberra, ACT, USA, and the University of Melbourne, Parkville, VIC, Australia.
His current research interests include security of cyberphysical systems, power systems, model reduction, and fundamental limitations in control.
Dr. Sandberg received the Best Student Paper Award from the IEEE Conference on Decision and Control in 2004, an Ingvar Carlsson Award
from the Swedish Foundation for Strategic Research in 2007, and Consolidator Grant from the Swedish Research Council in 2016.
He has served on the editorial boards of IEEE Transactions on Automatic Control and the IFAC Journal Automatica.
\end{IEEEbiography}

\end{document}